\documentclass[manuscript,screen,acmsmall]{acmart}
\AtBeginDocument{%
  }

\setcopyright{acmlicensed}
\copyrightyear{2018}
\acmYear{2018}
\acmDOI{XXXXXXX.XXXXXXX}
\acmISBN{978-1-4503-XXXX-X/2018/06}

\usepackage{graphicx}  
\usepackage{amsmath}   
\usepackage{multirow}
\usepackage{booktabs}
\usepackage{array}
\usepackage{subcaption}  
\usepackage{booktabs}
\usepackage{enumitem} 

\definecolor{bittersweet}{rgb}{0.3, 0.75, 0.3}
\newcommand{\zekun}[1]{\textcolor{black}{#1}}

\begin{document}

\title{How AI Responses Shape User Beliefs: The Effects of Information Detail and Confidence on Belief Strength and Stance}

\author{Zekun Wu}
\email{wuzekun@cs.uni-saarland.de}
\orcid{0000-0002-5233-2352}
\affiliation{%
  \institution{Saarland University}
  \streetaddress{Campus, 66123 Saarbrücken}
  \city{Saarbrücken}
  \state{Saarland}
  \country{Germany}
}

\author{Mayank Jobanputra}
\affiliation{%
  \institution{Saarland University}
  \streetaddress{Campus, 66123 Saarbrücken}
  \city{Saarbrücken}
  \state{Saarland}
  \country{Germany}
}
\email{jobanputra@cs.uni-saarland.de}

\author{Vera Demberg}
\affiliation{%
  \institution{Saarland University}
  \streetaddress{Campus, 66123 Saarbrücken}
  \city{Saarbrücken}
  \state{Saarland}
  \country{Germany}
}
\email{vera@coli.uni-saarland.de}

\author{Jessica Hullman}
\affiliation{%
  \institution{Northwestern University, Department of Computer Science}
  \streetaddress{2133 Sheridan Rd}
  \city{Evanston}
  \state{Illinois}
  \country{United States}
  \postcode{60208}
}
\email{jessica.hullman@northwestern.edu}

\author{Anna Maria Feit}
\affiliation{%
  \institution{Saarland University, Saarland Informatics Campus}
  \streetaddress{E1 7, 66123 Saarbrücken}
  \city{Saarbrücken}
  \state{Saarland}
  \country{Germany}
}
\email{feit@cs.uni-saarland.de}

\begin{abstract} \zekun{The growing use of AI-generated responses in everyday tools raises concern about how subtle features such as supporting detail or tone of confidence may shape people’s beliefs. To understand} this, we conducted a pre-registered online experiment (N=304) investigating how the detail and confidence of AI-generated responses influence belief change. \zekun{We introduce an analysis framework with two targeted measures: \emph{belief switch} and \emph{belief shift}. These distinguish between users changing their initial stance after AI input and the extent to which they adjust their conviction toward or away from the AI’s stance, thereby quantifying not only categorical changes but also more subtle, continuous adjustments in belief strength that indicate a reinforcement or weakening of existing beliefs.} Using this framework, we find that detailed responses with medium confidence are \zekun{associated with the largest overall} belief changes. Highly confident messages tend to belief shifts but induce less stance reversals. Our results also show that task type (fact-checking versus opinion evaluation), prior conviction, and perceived stance agreement further modulate the extent and direction of belief change. \zekun{These results illustrate how different properties of AI responses interact with user beliefs in often subtle but potentially consequential ways and raise practical as well as ethical considerations for the design of LLM-powered systems.}
\end{abstract}

\maketitle

\section{Introduction}

With the rapid advancement of large language models (LLMs), interactions between humans and artificial intelligence (AI) systems have become increasingly common. While earlier AI systems \cite{ferrucci2010building,croft2010search} often focused on retrieving and presenting facts, LLMs can generate fluent, persuasive, contextual responses that can influence how people perceive, interpret, and internalize this information. As these systems are integrated into everyday tools, a growing concern is how the framing of AI-generated messages can shape user beliefs and judgments, even when users remain in control of the final decision. Because LLMs communicate in natural language, they can subtly guide belief formation in ways that may not always be deliberate or consciously recognized by users. 


Prior research has shown that certain attributes of AI-generated responses—such as the amount of supporting detail and the tone of confidence—can affect how users interpret and respond to AI suggestions \cite{kim2025fostering, xu2025confronting,zhou2024relying}. For example, users are often more receptive to responses that include concrete evidence or cite external sources~\cite{kim2025fostering}, and may be more inclined to accept statements delivered with assertive rather than uncertain language~\cite{xu2025confronting}. However, much of this work focuses on discrete decision outcomes—such as whether a user accepts or rejects a suggestion—offering limited insight into how AI responses influence the continuum of belief change. \zekun{Belief change refers to how people update their convictions when exposed to new input, whether by holding on to prior views\cite{anglin2019beliefs}, shifting toward misinformation\cite{danry2024deceptive}, or reducing entrenched beliefs after corrective dialogue\cite{costello2024durably}.} A user may reach the same final decision after viewing an AI response, but with greater or reduced confidence in that decision. However, the different factors that influence such a belief shift are not yet well understood. 
In this study, we ask: \textit{"How do the detail and confidence of AI-generated responses affect the stance and strength of users’ beliefs during interaction?"}

\zekun{The answer to this question gives us a deeper understanding of how the content and style of AI responses can shape users’ internal beliefs, which is crucial for designing useful and ethical AI support systems. At the same time, these opportunities come with substantial ethical risks. AI responses that appear persuasive may also amplify misinformation, reinforce biases, or subtly undermine user autonomy. Because such influence often occurs below the level of conscious awareness, there is a danger that LLMs could be deployed in ways that manipulate rather than support users’ judgment. Recognizing these risks is essential not only for interpreting the results of our study, but also for ensuring that the design of LLM-powered systems incorporates safeguards that prevent misuse.}

\zekun{Such subtle belief changes cannot be captured by common outcome measures like decision accuracy, task performance, or confidence and trust scores. To address this gap, we propose a comprehensive analysis framework with two complementary components. First, we introduce two targeted measures—belief switch and belief shift—that capture, respectively, whether users abandon their initial stance after AI input and the extent to which they adjust their conviction toward or away from the AI’s stance. These measures provide a systematic way to quantify both categorical changes in stance and more continuous adjustments in belief strength. Second, to better illustrate the variety of belief change behaviors, we define five descriptive categories: flip (users reverse their stance toward the AI), reinforcement (users strengthen their stance when the AI agrees), weakening (users become less certain when the AI disagrees), unchanged (users’ stance and conviction remain stable), and deviation (users move against the AI’s stance). Together, these measures and categories allow us to go beyond binary outcomes and gain a more nuanced picture of how AI influences human belief.}

\zekun{To answer our research question, we applied this framework in a pre-registered online experiment (N=304) and examined how variations in the \textit{information detail} (low vs. high) and \textit{confidence level} (low, medium, or high confidence) of AI-generated responses influence users’ beliefs. The study was grounded in two AI-assisted task scenarios that reflect everyday use: \textit{fact-checking} and \textit{opinion evaluation}. Fact-checking captures the frequent situation where people quickly look up information online—for example, verifying the accuracy of a news headline or a historical detail—and increasingly encounter AI-generated answers embedded directly in search engines. Opinion evaluation, in turn, reflects the growing reliance on AI-generated comments and responses in online discussions, Q\&A forums, and social platforms, where people weigh in on subjective statements such as whether a policy is fair or a product is trustworthy. In our study, participants assessed whether a factual statement was true or false in the fact-checking task, first independently and then after receiving an AI-generated answer to a related question. In the opinion evaluation task, participants decided whether they agreed or disagreed with a subjective statement, again making their judgment both before and after viewing an AI comment that either supported or refuted the statement. In each task, we recorded both the stance (true/false or agree/disagree) and the strength of participants' beliefs before and after exposure to the AI response. Stance change was used to determine belief switch, while the shift in belief strength toward or against the AI’s stance served as the measure for belief alignment. Beyond the response features of detail and confidence, we also explored how other factors—such as task type, users’ initial belief strength, and their perceived agreement with the AI—may influence belief change.}

Our study reveals several new empirical insights. First, we found that both the detail and confidence level of AI responses significantly influence user belief change:  AI messages with high information detail and a moderately confident tone are most effective in persuading users to shift their beliefs toward the AI’s stance or prompt a belief switch. Notably, overconfident messages also increase belief shift but are less effective at inducing a belief switch. Second, we found that beyond AI message framing, several contextual and user-specific factors also shape belief change. For example, task type (e.g., fact-checking with objective ground truth vs. subjective opinion evaluation), users’ initial belief strength, and their perception of whether the AI agrees with them all play critical roles in determining how users update their beliefs after encountering AI suggestions. Finally, \zekun{a surprisingly larger number of trials fell in the deviation category where the participants went against what the AI suggested in their final decision}. Our analysis shows that such deviations are often driven by misperception of the AI’s stance and are more likely to occur when the AI expresses low confidence, which might lead to the users not trusting the AI's output. 

In summary, our work makes the following contributions: 
\begin{enumerate}
    \item We propose a framework for classifying and measuring human belief change during interactions with AI-generated responses. This framework captures not only decision outcomes but also the direction and strength of belief change, including belief flips, reinforcement, weakening, maintenance, and deviation from the AI's stance.
    \item Through a controlled experiment across two task scenarios where users relied on AI suggestions, we systematically examined how belief change is influenced by multiple factors—including AI response features (information detail and confidence tone), user-specific attributes (initial belief strength and \zekun{user-perceived AI agreement}), and task type. Our analysis reveals how these factors differentially affect the direction and extent of user belief change.
    \item We offer an in-depth discussion of the implications of these findings for the design and evaluation of LLM-based systems. Our results demonstrate that the framing of AI responses can be manipulated to interact with contextual and user-level factors in ways that shape human belief—raising both opportunities and ethical considerations for future AI design.
\end{enumerate}

\section{Related work}

\subsection{LLMs’ Influence on Human Belief}

LLMs are increasingly integrated into everyday tools, assisting users with a wide range of tasks such as information retrieval \cite{quelle2024perils, yan2024knownet, sharma2024generative}, opinion formation \cite{costello2024durably, xu2024role}, and decision-making \cite{lu2024does, eigner2024determinants}. Unlike earlier AI systems that primarily offered structured outputs or classifications, LLMs produce fluent, natural language responses that can directly influence human cognition \cite{karinshak2023working, bach2024systematic, jung2024quantitative, xu2024role, sharma2024generative, lu2024does}. \zekun{For instance, exposure to LLM-generated content can shape trust and adoption in academic and professional settings \cite{jung2024quantitative}. They also influence how people calibrate their confidence and reliance when evaluating misinformation \cite{xu2024role}. In persuasive contexts, LLM messages have been shown to shift attitudes, such as in public health communication \cite{karinshak2023working}. More broadly, they connect to known phenomena like automation bias, where users overweight machine advice \cite{lu2024does}, and can intensify selective exposure or echo chamber effects in information search \cite{sharma2024generative}.}

The ability of LLMs to generate coherent and persuasive language has sparked growing academic interest in how such content affects human beliefs and behavior. Earlier models like GPT-3, were already shown to produce persuasive outputs, such as pro-vaccination messages \cite{karinshak2023working} or propaganda~\cite{goldstein24} and persuasiveness has been shown to increase as models become larger and more capable~\cite{durmus2024persuasion}.
Besides persuasion, more recent studies examined the broader potential of LLMs to influence users at higher cognitive levels such as manipulation \cite{ienca2023artificial,borah2025persuasion} and deception \cite{costello2024durably,ji2023survey}. 

The potential for people to be influenced by LLM-generated content, even when the information is inaccurate or misleading, has raised serious ethical concerns about how these systems may shape human beliefs and behavior in problematic ways \cite{danry2024deceptive, borah2025persuasion, sabour2025human}. \zekun{For example, Sabour et al. \cite{sabour2025human} showed that users interacting with a manipulative AI agent, which was configured with hidden objectives to covertly steer them toward harmful options, were significantly more likely to make detrimental decisions compared to those assisted by a neutral agent that provided unbiased, user-benefiting advice.} \zekun{Danry et al. \cite{danry2025deceptive} showed that misleading explanations, defined as AI-generated justifications that present false claims as true or true claims as false, were more persuasive than accurate explanations, increasing belief in misinformation while undermining belief in accurate information.} 

On the other hand, LLMs also hold the potential to influence users in positive ways \cite{costello2024durably, kim2025fostering}. For instance,  \citet{costello2024durably} showed that an AI chatbot could help users reject conspiracy theories by providing tailored counterarguments and engaging them in personalized, in-depth conversations. \zekun{ \citet{schmitt2024role} found that in a fact-check task, crowdworkers’ accuracy improved significantly with AI support, especially with free-text explanations, reaching near journalist-level performance.}  

However, most existing studies that evaluate the influence of LLM-generated content on human belief or behavior tend to simplify this influence as a binary outcome, \zekun{such as whether a user switches their stance or accepts an AI recommendation\cite{lai2023towards,liu2024human,kim2025fostering,sabour2025human}} \zekun{, overlooking more subtle yet consequential changes in users’ beliefs after reviewing AI messages. Such changes, including reinforcing existing convictions, gradually weakening skepticism, or nudging beliefs in unintended directions, cannot be captured by discrete decision outcomes. Ignoring these dynamics risks underestimating the true extent of AI’s influence, obscuring cumulative shifts in attitude, polarization, or the reinforcement of biases that may occur even when overt decisions remain unchanged. To address this limitation, we propose a measurement framework that distinguishes between belief switch\zekun{—whether a user reverses their stance after seeing the AI response}—and belief shift, which captures the magnitude and direction of a user’s belief change relative to the AI’s stance.}

\subsection{Information Richness and Confidence Expression of AI Responses}

\zekun{Prior work shows that, even when an AI takes the same stance as a user, the specific content of the response, such as the use of explanations \cite{kim2025fostering,chen2023understanding,lee2023understanding,li2024utilizing}, the inclusion of citations \cite{ding2025citations,kim2025fostering}, or offering personalized suggestions \cite{salvi2025conversational}, can change a user's decisions. For example, Kim et al.  \cite{kim2025fostering} found that key elements such as response detail, internal consistency, and cited sources significantly influenced users’ fact-checking decisions and reliance on the AI’s answers. In addition to the content of AI responses, the way information is presented can also shape user reactions. One important factor is how confident the AI appears \cite{zhou2024relying, kim2024m, xu2025confronting}. \citet{zhou2024relying} showed that when AI responses included uncertainty cues, such as “I’m not sure”, users were more likely to verify the information themselves rather than rely on the AI’s suggestion.} Similarly, 
\zekun{Kim et al. \cite{kim2024m} found that expressing uncertainty in the first person (e.g., “I’m not sure, but...”), helps to reduce the user's overreliance on incorrect AI suggestions. In addition, \citet{xu2025confronting} found that appropriately expressed uncertainty led to higher user trust, satisfaction, and task performance compared to low or overly confident expressions of uncertainty. }


\zekun{In reviewing prior work, we find that most studies evaluated these features in isolation and often within a single task domain, such as misinformation detection or fact-checking \cite{kim2025fostering, xu2024role}, or persuasion in health communication \cite{karinshak2023working}. These studies typically focus on binary decision outcomes, for example, whether a user accepts or rejects an AI suggestion, without examining more gradual changes in belief strength. Moreover, they seldom compare domains with qualitatively different dynamics: fact-checking tasks involve verifiable ground truths that anchor decisions, whereas opinion evaluation tasks are inherently subjective with no single correct answer. In this work, we consider detail and confidence together and examine their effects across both objective (fact-checking) and subjective (opinion evaluation) contexts to offer a more integrated view of how content and tone shape user judgments.}

\section{Methods}

Our work aims to investigate how the information detail and confidence of AI-generated responses \zekun{shape both the stance people take and the strength with which they hold their beliefs} across two types of tasks: verifying factual claims (objective task) and evaluating open-ended opinions (subjective task). We conducted a controlled, pre-registered online experiment using Prolific, targeting U.S.-based participants with diverse demographic backgrounds, all of whom reported prior exposure to AI systems. 
In the following we first describe the task and design of the experiment. We then define a set of belief measures,  which allow a nuanced analysis of how users change their belief after seeing an AI message regarding both the magnitude and direction of their belief change in relation to the AI's stance.

\subsection{Experimental Task}

\begin{figure}[htbp]
  \centering

  \begin{minipage}{\textwidth}
    \centering
    \begin{subfigure}[t]{0.48\textwidth}
      \includegraphics[width=\textwidth]{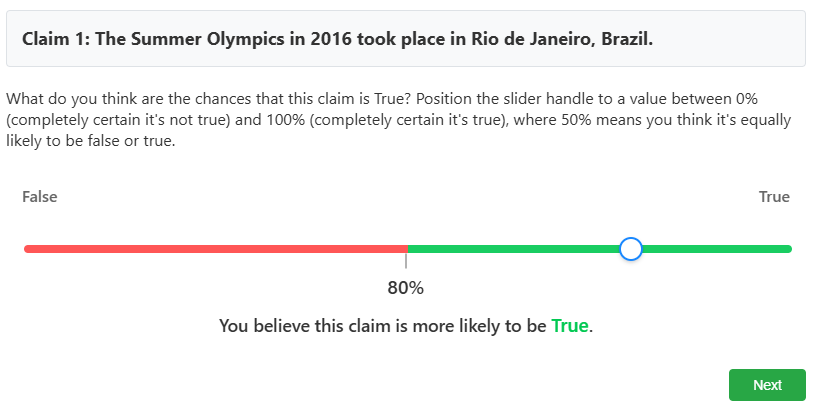}
    \end{subfigure}\hfill
    \begin{subfigure}[t]{0.48\textwidth}
      \includegraphics[width=\textwidth]{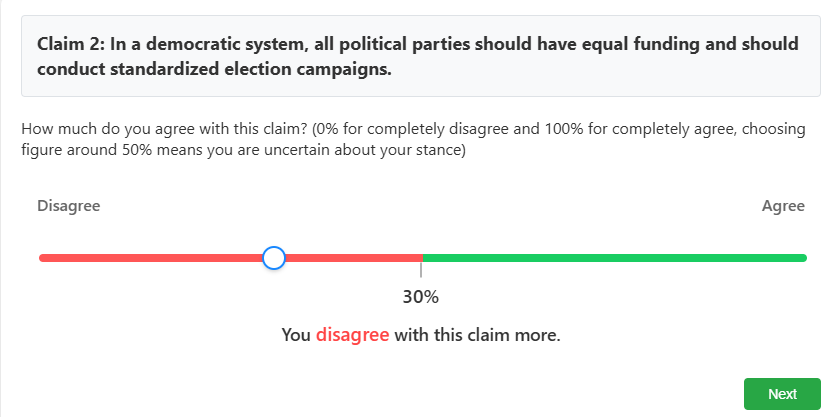}
    \end{subfigure}

    \subcaption*{\textbf{Step 1.} Initial belief (left: fact-checking; right: opinion evaluation)}
  \end{minipage}

  \vspace{0.9em}

  \begin{minipage}{\textwidth}
    \centering
    \begin{subfigure}[t]{0.48\textwidth}
      \includegraphics[width=\textwidth]{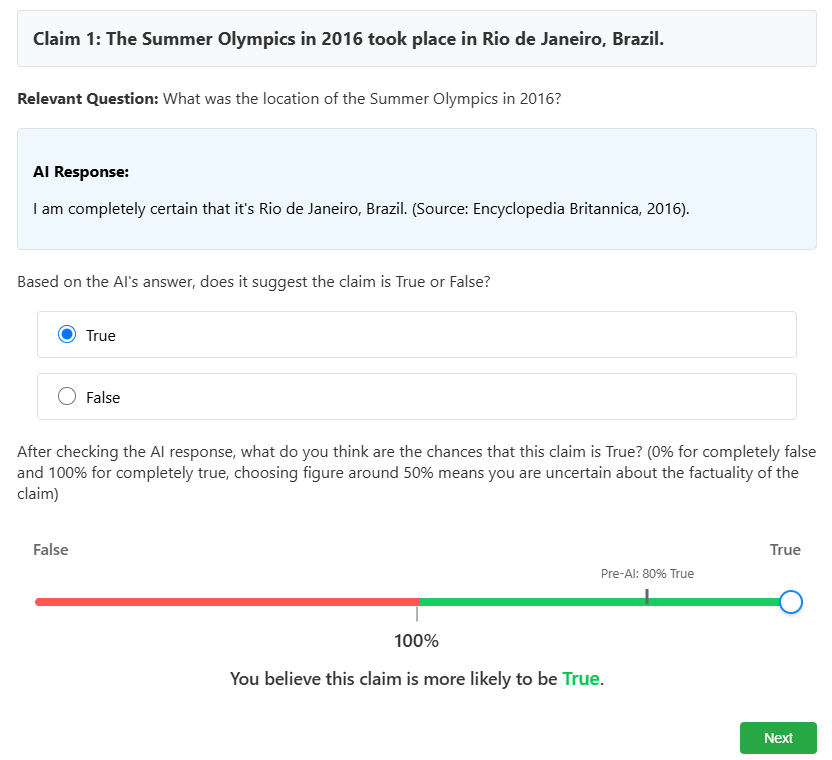}
    \end{subfigure}\hfill
    \begin{subfigure}[t]{0.48\textwidth}
      \includegraphics[width=\textwidth]{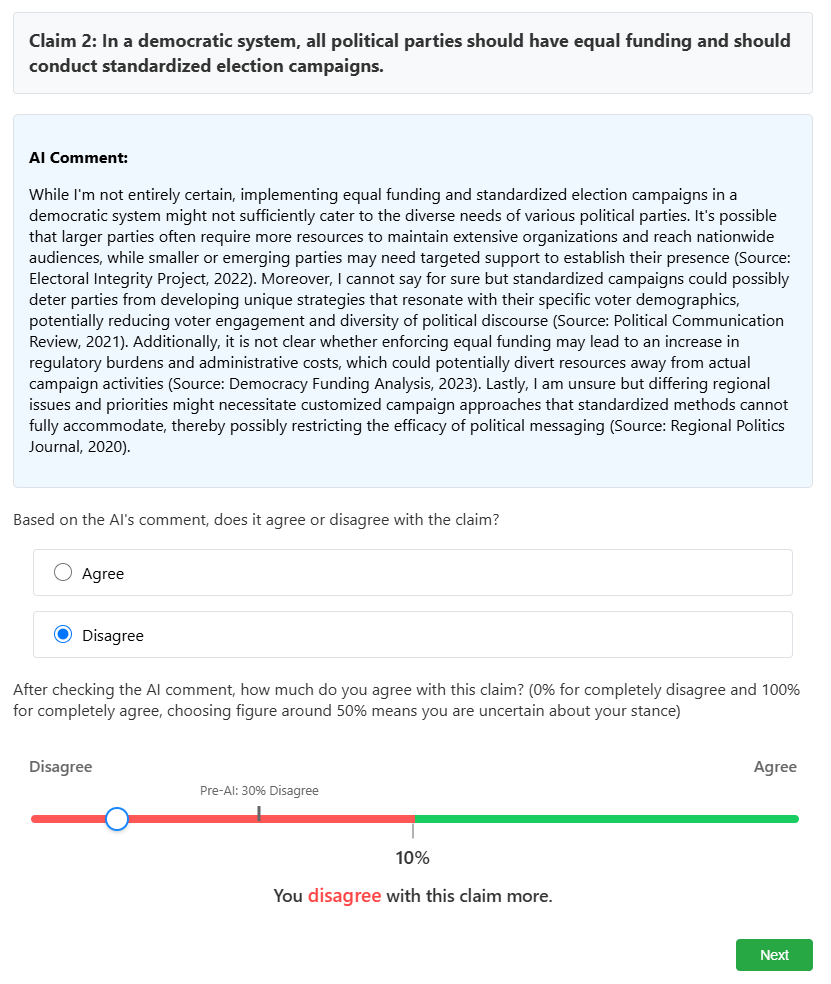}
    \end{subfigure}

    \subcaption*{\textbf{Step 2.} Re-evaluated belief after AI response (\zekun{left: fact-checking task with AI response in low detail and high confidence condition; right: opinion evaluation task with AI response in high detail and low confidence condition})}
  \end{minipage}

  \caption{Illustration of the two-step belief evaluation for the fact-checking (left) and opinion evaluation (right) task scenarios.}
  \label{fig:belief-eval}
\end{figure}

We designed the experiment around two task scenarios: fact-checking and opinion evaluation. In the fact-checking task (left column of \autoref{fig:belief-eval}), participants assumed the role of a fact-checker, reviewing factual claims submitted by users to a public knowledge database. In the opinion evaluation task (right column of \autoref{fig:belief-eval}), participants were told to imagine that they were community members on a discussion board, assessing and responding to subjective statements and personal viewpoints.

In both tasks, participants first indicated the strength of their beliefs about each claim, that is the probability they assigned to it being true (fact-checking) or the extent of their agreement (opinion evaluation), by adjusting a slider on a scale from 0\% (certainly false/completely disagree) to 100\% (certainly true/completely agree). The slider moved in 1\% increments. Importantly, the slider could not be placed at the neutral midpoint of 50\% to encourage participants to take a stance.


In the second step, participants were presented with an AI-generated response and asked to reevaluate their beliefs based on the AI’s output. In the fact-checking task, the AI response directly addressed a verification question related to the claim. For example, as indicated in the second step for fact-checking task in \autoref{fig:belief-eval}, the AI answers the question “What was the location of the Summer Olympics in 2016?”, providing information to assess the claim: “I believe it might be Rio de Janeiro, Brazil (Source: Encyclopedia Britannica, 2016).”
In the opinion evaluation task, the AI offered a supporting or opposing comment on the statement. For example, in the second step for opinion evaluation in \autoref{fig:belief-eval},  the AI disagrees with the claim and provides a detailed explanation while expressing strong confidence. The process for generating these AI responses, including the manipulation of information detail and confidence level, is described in the following section.
After viewing the AI response, participants were asked to indicate their belief again using the slider. Additionally, they completed a simple comprehension check on the same page, selecting either true/false (for fact-checking) or agree/disagree (for opinion evaluation) to indicate their interpretation of the AI’s stance.

\subsection{Experimental Design}

Our study employed a 2 (Information Detail: High vs. Low) $\times$ 3 (Confidence Level: Low, Medium, High) within-subjects factorial design to investigate how the content and tone of AI-generated responses affect user beliefs. \zekun{A within-subjects setup was chosen to directly compare framing effects across conditions for the same participants, thereby reducing the influence of individual differences in prior beliefs, conviction strength, and baseline trust in AI.} Each participant engaged in two types of tasks--fact-checking and opinion evaluation--involving exposure to AI responses that varied systematically across the six experimental conditions. For each task type, participants completed 12 trials, encountering two AI responses per condition and task type. The order of task trials was randomized across conditions and tasks (we did not block the trials by type). \zekun{In the opinion evaluation task, the AI response was equally likely (50\%) to support or oppose the participant’s initial belief, and supportive and opposing responses were alternated to ensure a balanced distribution of agreement across trials.}  In the fact-checking task, the AI always provided the correct answer, although participants were not informed of this. \zekun{This design choice allowed us to isolate the effects of factors (information detail and confidence level) on belief change without introducing the confounding factor of AI accuracy. In our study, this resulted in 63.3\% of trials where the AI response supported the users' initial stance (see results below), slightly more than in the Opinion Evaluation task.  While we acknowledge that LLMs can produce hallucinations and false information, our study deliberately focused on how framing alone shapes users’ belief stance and strength. }  Participants did not receive any feedback about their choices in either task.

\subsubsection{Materials}

For the fact-checking task, we sampled 120 factual claims along with their corresponding information-seeking questions from the FAViQ dataset \cite{park2021faviq}, focusing on the regular (non-opinionated) subset. These examples originate from real user search behavior and feature clearly verifiable answers. We manually reviewed all selected claims to ensure they were current, unambiguous, and not necessarily common knowledge, making it more likely that participants would attend to the AI-generated response \cite{kim2025fostering}. In addition to the claims and questions, we also collected the verified ground-truth answers from the dataset to serve as the basis for generating correct AI responses.

For the opinion evaluation task, we selected 120 subjective statements from the ChangeMyView dataset \cite{tan2016winning, atkinson+srinivasan+tan:19}, which contains controversial opinion posts collected from Reddit. Each statement was manually checked to ensure it was suitable for study---i.e., not offensive, context-independent, and reasonably open to both agreement and disagreement---so that AI responses could take either stance in a natural way.

\subsubsection{AI Response Generation}

\begin{table}[t]
\caption{Features for information detail and confidence level in AI Responses}
\label{tab:feature-definitions}
\footnotesize
\centering
\begin{tabular}{@{}m{1.5cm}m{2.5cm}m{4.0cm}m{5cm}@{}}
\toprule
\textbf{Feature Group} & \textbf{Feature} & \textbf{Definition} & \textbf{Calculation} \\
\midrule

\multirow{7}{=}{Information Detail} 
& Number of Citations & 
The number of external sources referenced in the response. & 
\textbf{Count} occurrences of citation markers (e.g., brackets ``[1]'', ``according to…'', named sources like ``Politifact'', ``EPA Report''). \\

& Number of Key Facts & 
The number of independent factual statements included in the response. & 
\textbf{Count} distinct factual claims (e.g., “the Earth revolves around the Sun.” and the Moon affects tides.” are counted as two separate facts).\\

& Number of Statistics & 
Whether the response includes numerical evidence to support claims. & 
\textbf{Count} occurrences of statistics such as percentages and ratio (e.g., 23\%, 5 out of 10, etc.). \\

\midrule

\multirow{4}{=}{Confidence Level} 
& Hedging (Uncertainty Markers) & 
Words or phrases that indicate uncertainty or possibility rather than certainty. & 
Classify as \textbf{Present} if hedging words (e.g., ``might'', ``possibly'', ``could'') appear; otherwise \textbf{Absent}. \\

& Definitive Statements & 
Words or phrases that indicate strong confidence in a claim. & 
Classify as \textbf{Present} if strong assertion markers (e.g., ``clearly'', ``proven'', ``without a doubt'') appear; otherwise \textbf{Absent}. \\

\bottomrule
\end{tabular}
\end{table}


\begin{table}[t]
\caption{Experiment conditions by information detail and confidence level}
\label{tab:exp-cond}
\footnotesize
\centering
\begin{tabular}{@{}c >{\centering\arraybackslash}p{1.5cm} >{\centering\arraybackslash}p{1.5cm} 
>{\centering\arraybackslash}p{1.6cm} >{\centering\arraybackslash}p{1.6cm} 
>{\centering\arraybackslash}p{1.5cm} >{\centering\arraybackslash}p{1.5cm} 
>{\centering\arraybackslash}p{1.5cm}@{}}
\toprule
\textbf{Cond.} & \textbf{Detail Level} & \textbf{Confidence Level} & 
\textbf{\# Citations} & \textbf{\# Key Facts} & 
\textbf{\# of Stats} & \textbf{Hedging} & \textbf{Definitive Statements} \\
\midrule

1 & Low  & Low & 1   & 1   & 0–1  & Present & Absent \\
2 & Low  & Medium  & 1   & 1   & 0–1  & Absent  & Absent \\
3 & Low  & High     & 1   & 1   & 0–1  & Absent  & Present \\

4 & High & Low & 2+  & 2+  & 2+   & Present & Absent \\
5 & High & Medium  & 2+  & 2+  & 2+   & Absent  & Absent \\
6 & High & High     & 2+  & 2+  & 2+   & Absent  & Present \\

\bottomrule
\end{tabular}
\end{table}

Notably, the AI responses generated by the LLM in our experiment systematically vary along two key dimensions: information detail and confidence level, as defined in \autoref{tab:feature-definitions}. Our study includes six experimental conditions resulting from the combination of two levels of information detail (low and high) and three levels of confidence (low, medium, and high), as summarized in \autoref{tab:exp-cond}.

To manipulate the information detail, we first constructed a base prompt containing only the claim, which produced a response aligned with the low-detail condition—featuring minimal citations, factual content, and numerical evidence—while maintaining a neutral tone with medium confidence. \zekun{In this way, we contrasted low-confidence responses (with hedging), high-confidence responses (with definitive markers), and medium-confidence responses (neutral tone without either) to test how these stylistic cues shape users’ belief adjustments.} We then created high-detail responses by supplying this initial output back to the LLM, along with additional factual input, to encourage the generation of responses with increased counts of the detail features listed in \autoref{tab:feature-definitions}, such as multiple citations or statistics. Notably, at this stage, both the low- and high-detail responses were generated with medium confidence. We also observed that, by default, LLM responses lacked uncertainty modifiers (e.g., “might,” “possibly”), which naturally aligned with our goal of using these outputs to represent the neutral confidence condition.


To manipulate the confidence level, we identified common hedging and definitive language patterns based on prior work \cite{xu2025confronting}. Hedging expressions included phrases such as \textit{"I’m not sure,"} \textit{"I cannot provide a definitive answer,"} and \textit{"It is not clear."} Definitive expressions included statements like \textit{"I am confident,"} \textit{"Undoubtedly,"} and \textit{"I can confidently say"}. We integrated the identified hedging or assertive expressions into new prompts, and combined them with the medium-confidence responses generated in the previous step. Then these updated prompts were submitted to the LLM to produce both low- and high-confidence variants, without altering the underlying content or level of detail.

For the fact-checking task, each claim was paired with six AI responses—one for each condition. For the opinion evaluation task, where both agreement and disagreement are relevant, each opinion prompt was paired with twelve responses (six supporting and six refuting). All prompt templates and AI-generated response examples used in our study are available in the supplementary materials.

\subsection{Study Procedure}
\label{sec:study-procedure}

Upon agreeing to participate in the study, participants first completed a brief demographic survey, providing information about their gender, age, educational background, familiarity with AI systems, and general trust in AI. After the survey, they watched a short introductory video that explained the study context and outlined the experimental procedure. Specifically, the video explained the hypothetical role of the participant as fact-checker or member of a discussion board and that the messages they received to support them in their task came from an AI-based system. 


In each trial, participants were first presented with a randomly selected claim, either a factual statement that required verification or a subjective statement that expressed an opinion, from the dataset we constructed. They reported their initial belief position (true/false for fact-checking, agree/disagree for opinion evaluation) and their level of certainty using the slider shown in \autoref{fig:belief-eval}. They were then shown an AI-generated response, where the tone (confidence level) and information richness (detail level) were systematically manipulated based on the experimental condition. In the fact-checking task, the AI provided a direct factual answer; in the opinion evaluation task, it offered a supportive or opposing comment. 

After viewing the AI response, participants completed a comprehension check, indicating how they perceived the AI’s stance with a binary choice (true/false for fact-checking or agree/disagree for opinion evaluation). Participants then indicated their belief again using the slider, with their pre-AI belief explicitly marked on the scale for reference. Finally, they answered a question assessing their perceived influence of the AI on their final decision.

Each participant completed 25 trials, one of which is an attention check, where the AI explicitly instructed participants to select a specific answer; participants were informed about the presence of such trials in advance. After completing all trials, participants responded to an open-ended behavioral question, describing which specific aspects of the AI response influenced their final judgment.


\subsection{Analysis of Belief Change}
\label{sec:belief-change-analysis}
To examine how task type and AI-generated responses influence user beliefs, we defined a set of belief measures that capture the user's agreement with the AI, as well as the magnitude of their belief change after seeing the AI response. Additionally, we define five categories of belief change to provide a more nuanced assessment of the extent to which users changed their beliefs to align with the AI response or to deviate from it.
In the following, we first define these measures. We then formulate our hypotheses and describe how we analyzed our data.

\subsubsection{Pre-processing and Terminology}

The experiment interface elicited participants' beliefs for both the fact-check and opinion evaluation tasks using a 0-100 scale. 
We transformed the responses so that they were centred around zero by subtracting 50 from each value. This yielded a new belief scale on the interval [–50, 50]. 
Formally, we represent a participant’s \textit{belief} as a single scalar value \( x \in [-50, -1] \cup [1, 50] \) (recall the midpoint could not be selected by participants to encourage participants to take a stance). 
This value captures two aspects of the belief: \textit{belief stance} and \textit{belief strength}. 
The sign of \( x \) indicates the participant’s \textit{belief stance}, that is, whether they considered the claim to be more likely to be true or false (in the fact-checking task), or whether they tended to agree or disagree with the statement (in the opinion evaluation task). 
The absolute value \( |x| \) represents the \text{belief strength}, which reflects how strongly they leaned toward that stance.
We denote a participant's \textit{initial belief}, before viewing the AI response, by \( (x_i) \), and their \textit{post-AI belief} by \( x_p \). The stance of the AI response is a binary value denoted by \( a \in \{-1, 1\} \). In the following,  we use ``agree/disagree'' to describe whether the AI’s suggestion matched the participant’s initial belief stance, and “changed/maintained” to refer to whether the participant revised their belief after seeing the AI response. We describe whether the AI's suggested stance matched the participant’s belief stance as \textit{user-AI agreement}.

\subsubsection{Belief changes measures}
\label{belief_change_measure}
To capture how participants change their beliefs after viewing AI responses, we define two complementary measures. 

\textit{Belief switch}, is a categorical variable that captures whether the participant reverses their stance after seeing the AI response. It is defined as:

\[
\text{belief switch} = 
\begin{cases}
  1 & \text{if } x_p \cdot x_i < 0 \\
  0 & \text{otherwise}
\end{cases}
\]

\textit{Belief shift} is defined as a continuous variable that captures how much the participant's belief moved toward or away from the AI’s stance. It is defined as:

\[
\text{belief shift} = a \cdot (x_p - x_i)
\]
The \textit{belief shift} score ranges from –100 to +100, with a positive value indicating the magnitude of belief change towards the AI's stance and negative values indicating a belief change against the AI's stance. 

In addition to measuring belief change, we also collected participants’ self-reported perceptions of how much the AI response influenced their decision. After each trial, participants rated the \textit{perceived AI influence} on a scale from 0 to 100.

\subsubsection{Belief Change Category}


\zekun{While belief switch and belief shift provide quantitative measures of stance reversal and magnitude of change, they do not distinguish between qualitatively different belief trajectories, for instance, whether a user strengthens an already aligned stance, softens an opposing stance, or actively moves against the AI.} Based on the belief definition and measure that were formulated above, we then classify user belief changes in response to AI suggestions into five more interpretable categories: \textit{flip}, \textit{reinforcement}, \textit{weakening}, \textit{unchanged}, and \textit{deviation}. These categories are determined by analyzing the transition from the initial belief state \( x_i \) to the updated state \( x_p \), relative to the AI’s suggested position \( a \). As shown in \autoref{tab:belief-change-types}, the first three belief change types—flip, reinforcement, and weakening—reflect cases where users shift their belief towards the AI’s stance. Specifically: (1) \textit{flip} occurs when the user's stance changes to align with the AI’s stance; (2) \textit{reinforcement} occurs when the user maintains the same stance as the AI but increases their belief strength; and (3) \textit{weakening} occurs when a user maintains a difference stance compared to the AI reduces their conviction. \textit{Unchanged} refers to cases where both belief position and strength remain stable. Finally, \textit{deviation} captures changes in the participant's belief that go against the AI’s stance.



\begin{table}[h!]
\centering
\caption{Classification of Belief Change Based on User-AI Agreement and Belief Transition}
\label{tab:belief-change-types}
\resizebox{\columnwidth}{!}{%
\begin{tabular}{>{\centering\arraybackslash}m{2.0cm} 
                >{\centering\arraybackslash}m{3.2cm} 
                >{\centering\arraybackslash}m{4.6cm} 
                >{\centering\arraybackslash}m{3.2cm}
                >{\centering\arraybackslash}m{2.5cm}
                >{\centering\arraybackslash}m{3.5cm}}
\toprule
\multirow{2}{*}{\textbf{Category}} & 
\multirow{2}{*}{\textbf{User-AI Agreement}} & 
\multirow{2}{*}{\textbf{User Stance Change}} & 
\multirow{2}{*}{\textbf{Belief Change}} & 
\multicolumn{2}{c}{\textbf{Measure}} \\
\cmidrule(l){5-6}
& & & & \textbf{Switch} & \textbf{Shift} \\
\midrule
\textit{Flip} & Disagree (\( x_i \cdot a < 0 \))  
& Changed (\( x_p \cdot a > 0,\ x_p \cdot x_i < 0 \))     
& \( Either(\ast) \) 
& 1 
& \multirow{3}{*}{\( a \cdot (x_p - x_i) > 0 \)} \\
\cmidrule(l){1-5}
\textit{Reinforcement} & Agree (\( x_i \cdot a > 0 \))     
& Maintained (\( x_p \cdot a > 0,\ x_p \cdot x_i > 0 \))           
& Increase (\( x_p - x_i > 0 \)) 
& \multirow{3}{*}{0} 
&  \\
\cmidrule(l){1-4}
\textit{Weakening} & Disagree (\( x_i \cdot a < 0 \))  
& Maintained (\( x_p \cdot a < 0,\ x_p \cdot x_i > 0 \))        
& Decrease (\( x_p - x_i < 0 \)) 
&  
&  \\
\cmidrule(l){1-4} \cmidrule(l){6-6}
\textit{Unchanged} 
& \( Either(\ast) \)                        
& Maintained (\( x_p \cdot x_i > 0 \)) 
& Maintained (\( x_p - x_i = 0 \)) 
&  
& 0 \\
\cmidrule(l){1-6}
\textit{Deviation} 
& \multicolumn{3}{c}{All other cases} 
& \( Either(\ast) \) 
& \( a \cdot (x_p - x_i) < 0 \)  \\
\bottomrule
\end{tabular}%
}
\end{table}

\subsubsection{Hypotheses}
\label{subsec:hypotheses}
We formulate three hypotheses to test \zekun{\emph{how the detail and confidence of AI-generated responses affect the stance and strength of user's belief} and \emph{how this is further influenced by context.}
Specifically, we formulate the following hypotheses: }

\textbf{H1}: AI-generated responses will have a stronger impact on belief change and perceived AI influence when users evaluate factual claims compared to subjective opinions.

\textbf{H2}: Responses with high information detail will lead to greater belief change and higher perceived AI influence than low-detail responses.

\textbf{H3}: Responses expressed with higher confidence will lead to greater belief change and higher perceived AI influence than responses with lower confidence.


\subsubsection{Analysis Approach}
\label{subsec:modelling}
\zekun{To test our hypotheses in the \autoref{subsec:hypotheses},  we rely on generalized mixed-effects regression models. Specifically,  we focus on three dependent variables introduced in \autoref{belief_change_measure}: \textit{belief switch}, \textit{belief shift}, and \textit{perceived AI influence}. For the binary dependent variable (\textit{belief switch}) we used logistic mixed-effects model, and for continuous dependent variables (\textit{belief shift} and \textit{perceived AI influence}) we used a linear mixed-effects model}. 

\zekun{Because each participant evaluated multiple claims and multiple participants saw each claim, the data had a repeated-measures and crossed design. To account for this, we included random intercepts for both participants and claims. This specification captures individual- and claim-level variability in baseline responses. In line with our preregistration, the fixed-effects structure of the models included our manipulated variables—task type, information detail, and confidence level—together with covariates that plausibly influence belief change: initial belief strength, perceived agreement with the AI, and trial order. This combination allowed us to test the effects of experimental manipulations while controlling for participant-level and task-level differences.} The full model specification is detailed in \autoref{sec:belief_change_effects_test}

In addition to model-based hypothesis testing, we also conduct descriptive analyses of belief change outcomes, using the belief change taxonomy defined in \autoref{tab:belief-change-types}. This descriptive view—covering cases like flip, reinforcement, and deviation—helps contextualize user behavior before introducing statistical modeling and yields further insights to our research question.


\subsection{Participants}  
To determine the number of participants, we performed a simulation-based power analysis using a generalized linear mixed-effects model (GLMM) fit to a pilot dataset of 57 people. The model predicts the likelihood of a participant flipping their belief toward the AI suggestion based on confidence level, detail level, their interaction, and several covariates (task type, pre-AI certainty, trial number, and whether the participant initially agreed with the AI).
\zekun{For the power analysis, we overwrote the pilot effect sizes with target values: a positive main effect of confidence level (0.8), a positive main effect of detail level (0.6), and a negative interaction effect (-0.6). These values were chosen to represent plausible expected effects in the full study.}\zekun{We then generated simulated datasets by sampling belief-switch outcomes from the fitted GLMM, using the pilot-estimated fixed effects and random-effect variance components. We simulated datasets with up to 500 participants and conducted power analyses across several sample sizes (100, 200, 300, 400, and 500).} The results showed that at 300 participants, we would achieve above 80\% power to detect the main effects and interaction effects. Therefore, we set our target sample size to 300 participants.

Specifically, we recruited 320 U.S.-based adult participants through Prolific. Following our pre-registered exclusion criteria, we removed 16 participants: 6 for failing the attention check. Then we further exclude 10 for achieving less than 50\% accuracy on the fact-checking task. The final sample included 304 participants. Of these, 54.7\% self-identified as female and 44.7\% as male. The participants ranged in age from 18 to 83 years, with a median age of 37 years. The majority (84.7\%) of the participants held a bachelor's degree or higher. Most participants (97.5\%) reported having at least basic experience using LLMs in their daily lives, and 94.1\% indicated a moderate or higher level of trust in AI. Regardless of inclusion or exclusion in the final sample for analysis, we paid all participants \$5.60 as a base payment and an additional \$1.60 bonus based on their accuracy in the fact-checking task. The median duration of the study was 32 minutes, resulting in an average effective pay rate of at least \$13.51 per hour.

After conducting a pilot experiment, we pre-registered our study on OSF\footnote{\url{https://osf.io/8ju7v/?view_only=adacbd55ce0e40148b251ea72bef1ab4}}. The study was approved by our institutional review board (IRB).

\section{Results}


We organize the results into three parts. We begin with descriptive analyses that summarize general patterns in user beliefs before and after AI interaction, including changes in user-AI agreement, belief strength, and the distribution of belief change outcomes across predefined categories. We then turn to hypothesis testing, using mixed-effects regression models to evaluate how task type and features of the AI response influence belief change, as outlined in \autoref{subsec:hypotheses}. Finally, we present exploratory analyses focused on two phenomena observed in the data: misperception of the AI’s stance and belief deviation.

\subsection{Descriptive Analysis of Belief Change}

\subsubsection{User-AI Agreement and Belief Strength Change Across Tasks}

We began our analysis with descriptive summaries examining how participants' beliefs changed after receiving AI responses in the two task scenarios. To capture these changes, we used two measures introduced in \autoref{sec:belief-change-analysis}: \textit{user-AI agreement} and
\textit{belief strength}

\begin{figure}[htbp]
    \centering
    \begin{subfigure}{0.49\textwidth}
        \includegraphics[width=\linewidth]{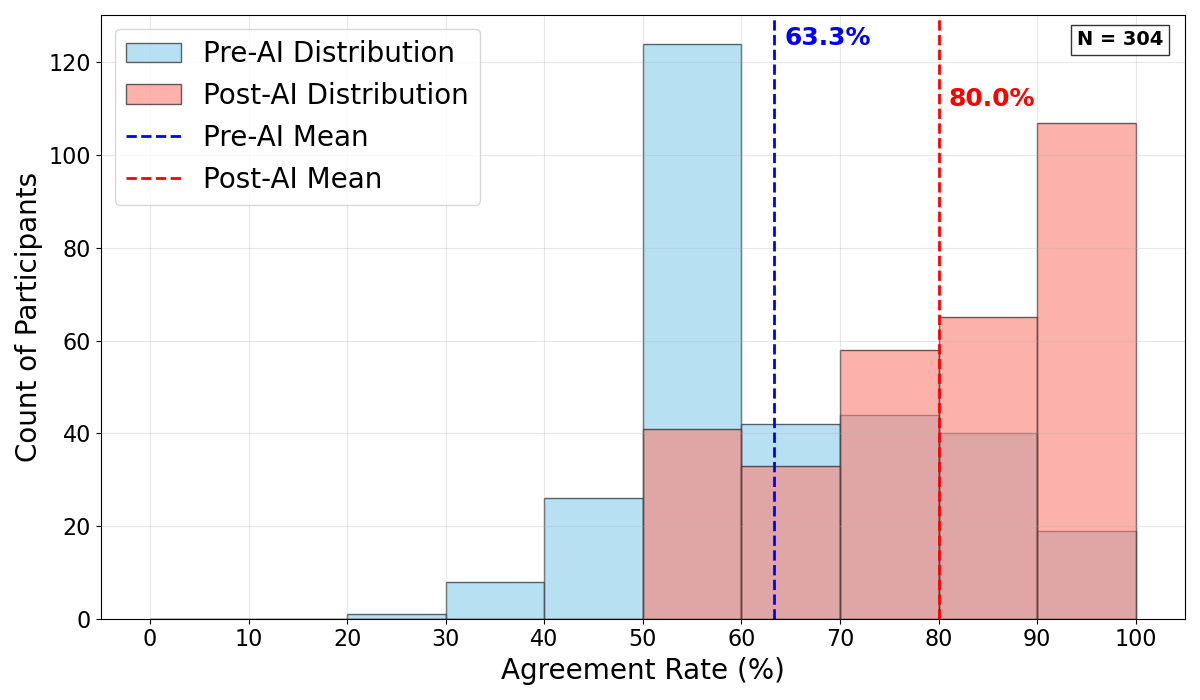}
        \caption{Fact-check}
        \label{fig:fact-check-agreement}
    \end{subfigure}
    \hfill
    \begin{subfigure}{0.49\textwidth}
        \includegraphics[width=\linewidth]{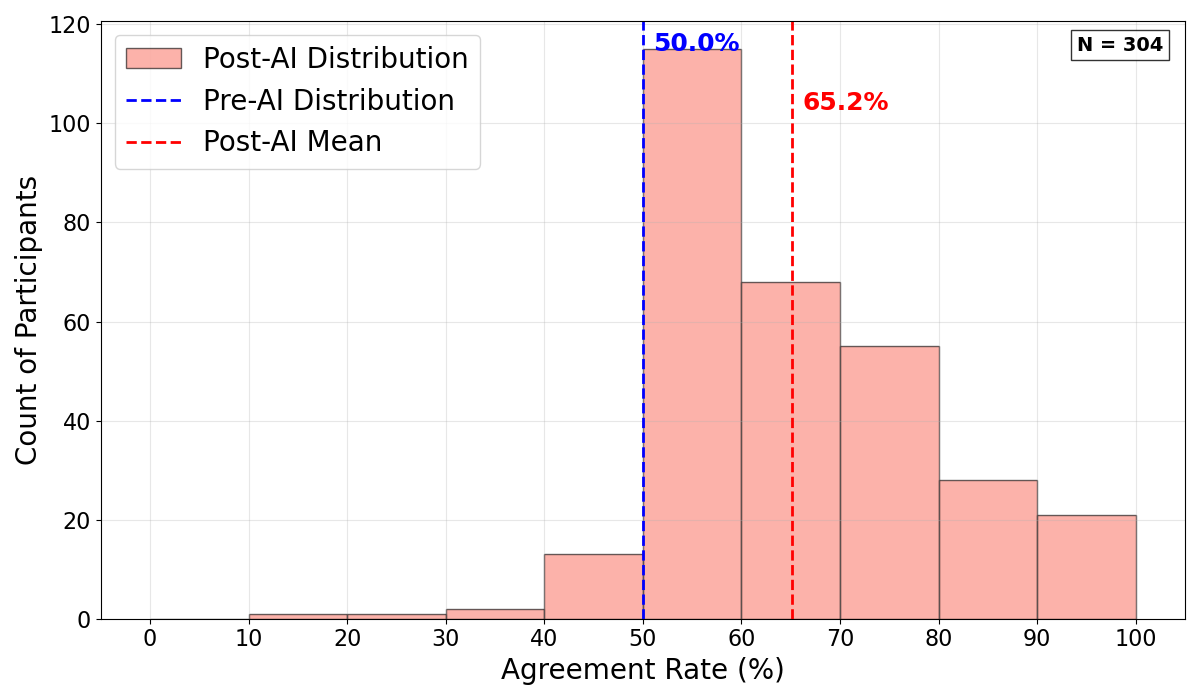}
        \caption{Opinion Evaluation}
        \label{fig:opinion-agreement}
    \end{subfigure}
    \caption{Rate of agreement between each participant's and AI's stance for the initial and post-AI beliefs in both tasks. Note that for the Fact-check task, agreement is the same as task accuracy, since the AI always provided correct evidence in support or against the claim. In the Opinion Evaluation task, Pre-AI alignment was controlled, since we randomly presented participants with opposing AI views for half of the trials.}
    \label{fig:pre_post_AI_agreement_task}
\end{figure}

As shown in \autoref{fig:pre_post_AI_agreement_task}, we computed the average agreement rate for each participant and visualized the distribution before and after AI exposure. In the fact-checking task---where the AI always provided the correct answer---the participant-level agreement rate (which also reflects accuracy) increased from an average of 63.3\% to 80.0\%. In the opinion evaluation task, agreement increased from 50.0\% (which was controlled by the study design) to 65.2\%. While the average gain was similar across tasks (16.7 percentage points in fact-checking vs. 15.2 in opinion evaluation), the higher absolute agreement rates in the fact-checking task suggest that users aligned more strongly with the AI when there was an objective ground truth to anchor their beliefs.

\begin{figure}[htbp]
    \centering
    \begin{subfigure}{0.49\textwidth}
        \includegraphics[width=\linewidth]{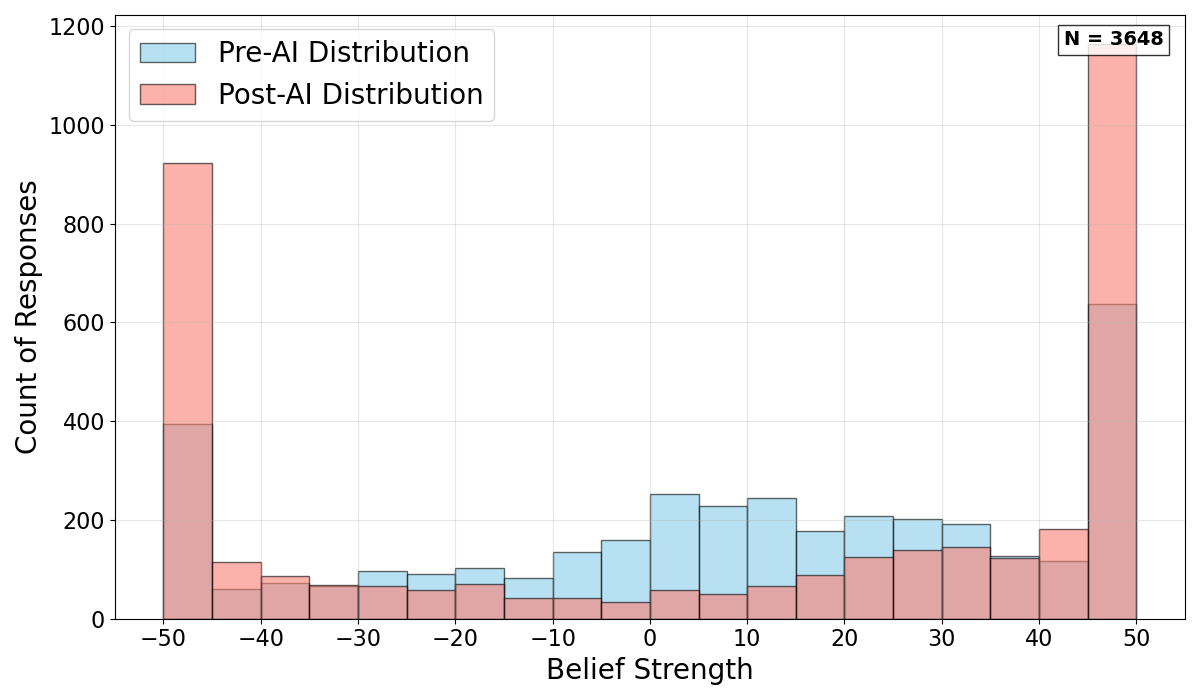}
        \caption{Fact-check}
        \label{fig:belief_strength_fact}
    \end{subfigure}
    \hfill
    \begin{subfigure}{0.49\textwidth}
        \includegraphics[width=\linewidth]{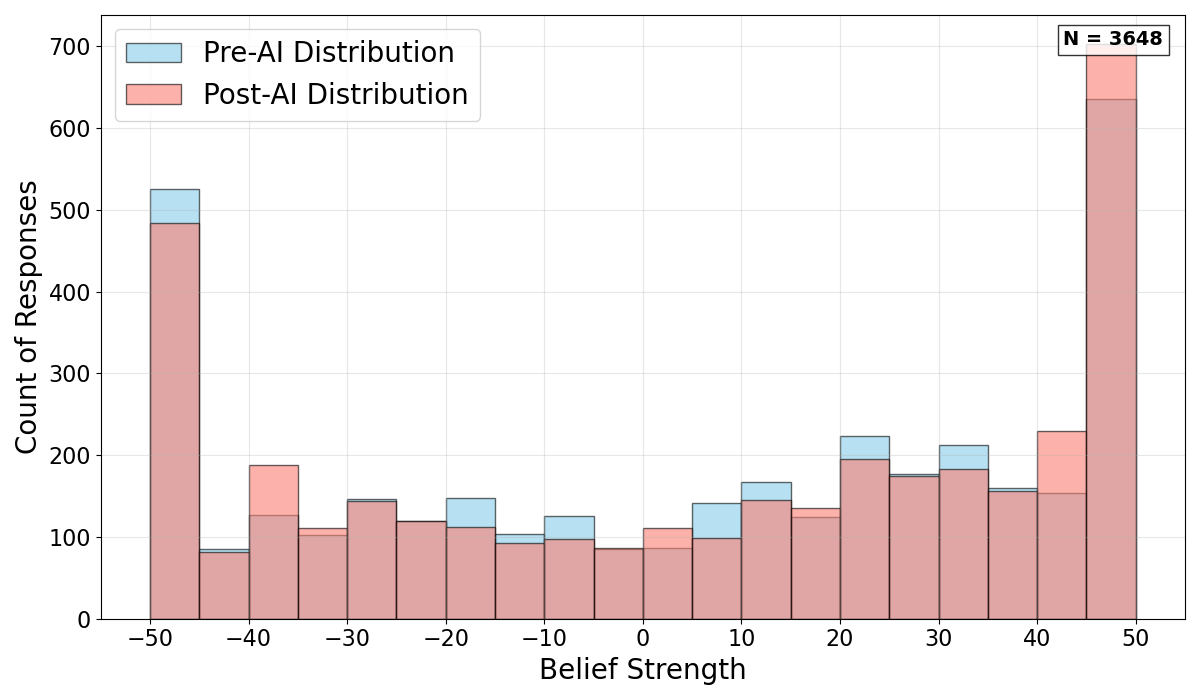}
        \caption{Opinion Evaluation}
        \label{fig:belief_strength_opinion}
    \end{subfigure}
    \caption{Changes in the strength of participants' beliefs pre- and post-AI by task type. \zekun{For the Fact-Check task we see that participants' belief strength increases after seeing the AI's response. This effect is less pronounced in the Opinion Evaluation task. }}
    \label{fig:pre_post_AI_belief_strength_tasks}
\end{figure}

We also analyzed how participants' belief strength changed after receiving the AI response. As shown in \autoref{fig:pre_post_AI_belief_strength_tasks}, belief strength became more polarized in the fact-checking task (\autoref{fig:belief_strength_fact}), with participants' reported beliefs clustering more strongly at the extremes of the scale after they viewed the AI suggestion. \zekun{This likely occurs because fact-checking provides an objective anchor: when the AI aligns with one’s initial stance, it reinforces it, and when it contradicts it with verifiable evidence, participants often update in that direction with higher certainty}

\subsubsection{Belief Change Outcomes by Actual vs. Perceived AI Stance}
\label{subsec:belief_change_ai_position}

\begin{figure}[h]
\centering
\includegraphics[width=5.0in]{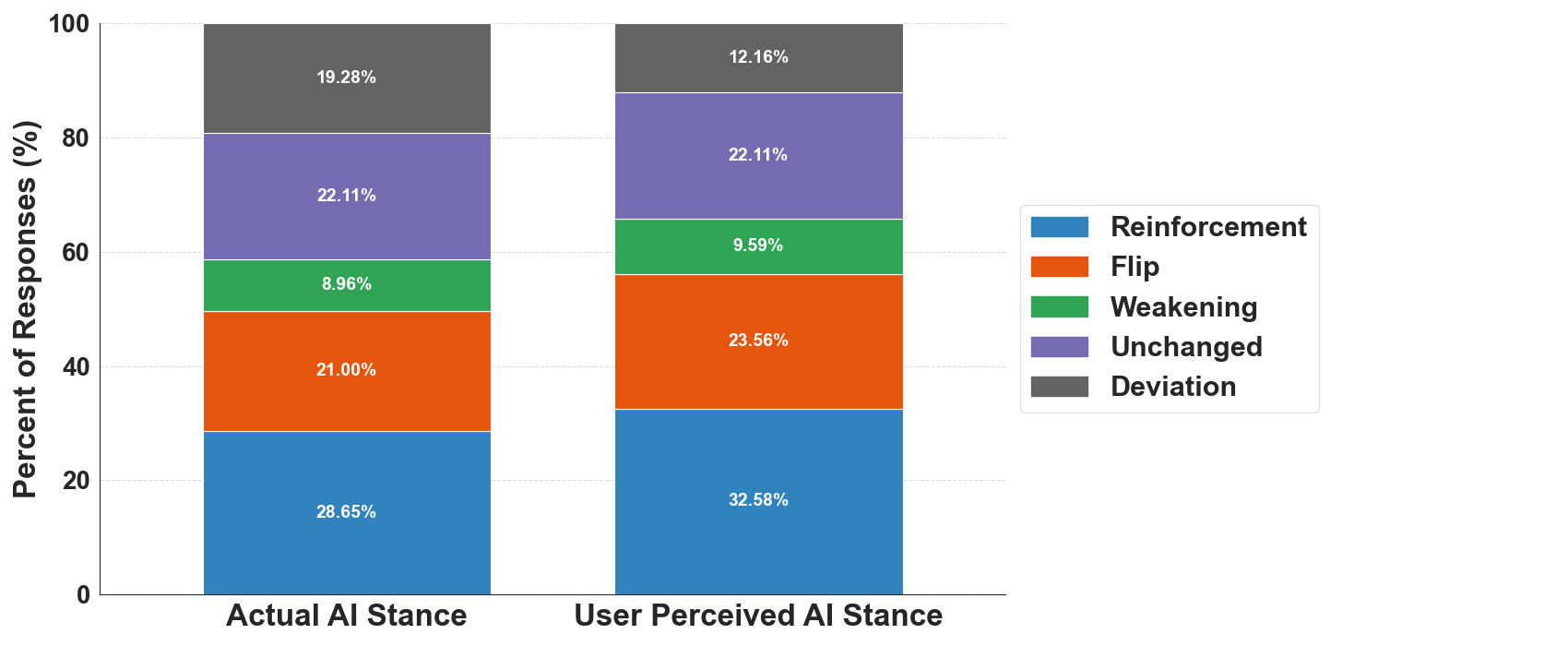}
\caption{Distribution of User Responses across the Belief Change Categories based on Actual AI Stance and User Perceived AI Stance}
\label{fig:ai_influence}
\end{figure}

Next, we examine the distribution of belief change categories based on the definitions in \autoref{tab:belief-change-types}. As shown in the left stacked bar chart in \autoref{fig:ai_influence}, using the actual AI stance as the reference, we observed the following belief change distribution across all responses. In 21.0\% of cases, participants flipped their belief stance to align with the AI’s opposing stance. Another 28.7\% of responses showed reinforcement, where participants maintained and strengthened their initial belief when the AI confirmed their stance. We also observed 9.0\% of responses where participants weakened their belief without fully reversing it, despite the AI suggesting an opposing stance. In 22.1\% of cases, participants did not change their belief at all, regardless of what the AI suggested. Notably, 19.3\% of cases involved deviation, where participants changed their stance in the opposite direction of the AI’s stance—for example, switching their belief even when the AI supported their original stance.

To understand how belief change distribution might differ based on how participants interpreted the AI’s stance, we reclassified the responses using the perceived AI stance, as shown in the right stacked bar chart in \autoref{fig:ai_influence}. When accounting for participants’ perceptions, deviation cases percents dropped to 12.2\%, while flip, reinforcement and weakening cases' percents increased to 23.6\%, 32.6\%, and 9.6\%  respectively. This comparison reveals a noticeable discrepancy between the AI’s actual stance and how participants perceived it, highlighting the role of subjective interpretation in belief change. We further examine this misperception and its implications in \autoref{sec:deviation}.

\subsection{Model-Based Analysis of Belief Change}
\label{sec:belief_change_effects_test}
As mentioned in the \autoref{subsec:hypotheses}, we hypothesized that the features of LLM responses that we manipulated — the information detail and the confidence level — would affect user belief before and after seeing the AI suggestion. Besides these two independent variables, our regression models also included several control variables that are likely to influence belief change:

\begin{itemize}
 \item \textbf{Task Type:} A categorical variable indicating the type of task participants completed—fact-checking or opinion evaluation. This variable was dummy-coded in the regression models, with fact-checking used as the reference level.

  \item \textbf{Initial Belief Strength: }The absolute value of the participant's initial belief rating, indicating how confident they were in their stance before viewing the AI response.

  \item \textbf{User Perceived AI Agreement:} Whether the participant believed the AI’s suggestion agreed with their initial belief stance. We include this factor because we observed differences between actual and perceived user-AI agreement. Detailed results on how misperceiving AI stance affects belief change are shown in \autoref{sec:deviation}.

  \item \textbf{Trial Number:} The index of the trial in the sequence of tasks completed during the experiment.
\end{itemize}

\zekun{To test our hypotheses (H1–H3) listed in \autoref{subsec:hypotheses}, we used generalized mixed-effects regression models, with the choice of logistic or linear depending on the belief change measure. Each of the three belief measures introduced in Subsection 3.4.2—belief switch, belief shift, and perceived AI influence—served as a dependent variable (denoted as BM below). The predictors included our manipulated factors, Confidence Level and Detail Level (and their interaction), as well as control variables identified earlier (Task Type, Initial Belief Strength, Perceived AI Agreement, and Trial Number). To account for repeated measures, we included random intercepts for both participants and claims. The resulting model specification was: \( \text{BM} \sim \text{Confidence\_Level} \times \text{Detail\_Level} + \text{Task\_Type} + \text{Initial\_Belief\_Strength} + \text{Perceived\_AI\_Agreement} + \text{Trial\_Number} + (1 \mid \text{Participant\_ID}) + (1 \mid \text{Claim\_ID}) \), where \texttt{Detail\_Level} uses the low-detail condition as the reference level, and \texttt{Confidence\_Level} uses the low confidence condition as the reference level. We report and interpret our results using both p-values for statistical significance and effect sizes to quantify the main difference between the groups.}


Study materials, including the collected data and a notebook of analyses (in R), are available as supplementary material to this submission.

\subsubsection{Belief Switch}
\label{subsec:belief_switch}
\zekun{Recall that a belief switch indicates whether users changed their initial belief stance after receiving AI input. }


\zekun{Our model-based analysis described above} found a significant main effect of task on belief switch ( $\beta = -1.52$, $SE = .13$, $p < .001$), \zekun{indicating that participants more often changed their stance after seeing the AI's response in the Fact-Check task compared to the Opinion Evaluation}. 
Regarding the AI response features, we observed a significant main effect of information detail ($\beta = .87$, $SE = .18$, $p < .001$).  Participants were more likely to switch their belief stance in response to AI messages with high detail compared to responses with low detail. The confidence level of the AI response also played a role: compared to low-confidence messages, AI responses expressed with medium confidence were more likely to trigger a belief switch ($\beta = .76$, $SE = .18$, $p < .001$). This effect was not observed for highly confident messages, suggesting that overconfidence may reduce persuasive impact, which is consistent with previous findings \cite{xu2025confronting}. No significant interaction effect was found between information detail and confidence level. 

\begin{figure}[htbp]
    \centering
    \begin{subfigure}{0.32\textwidth}
        \includegraphics[width=\linewidth]{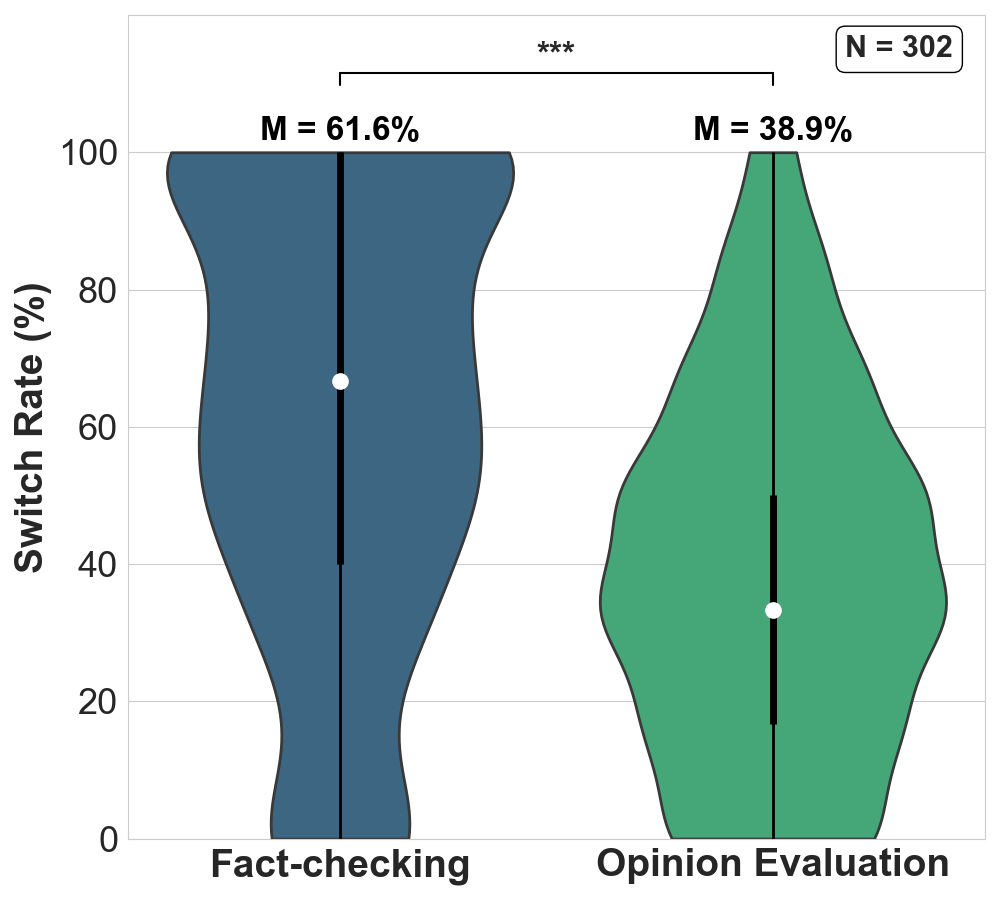}
        \caption{Task Type}
        \label{fig:switch_rate_task}
    \end{subfigure}
    \hfill
    \begin{subfigure}{0.32\textwidth}
        \includegraphics[width=\linewidth]{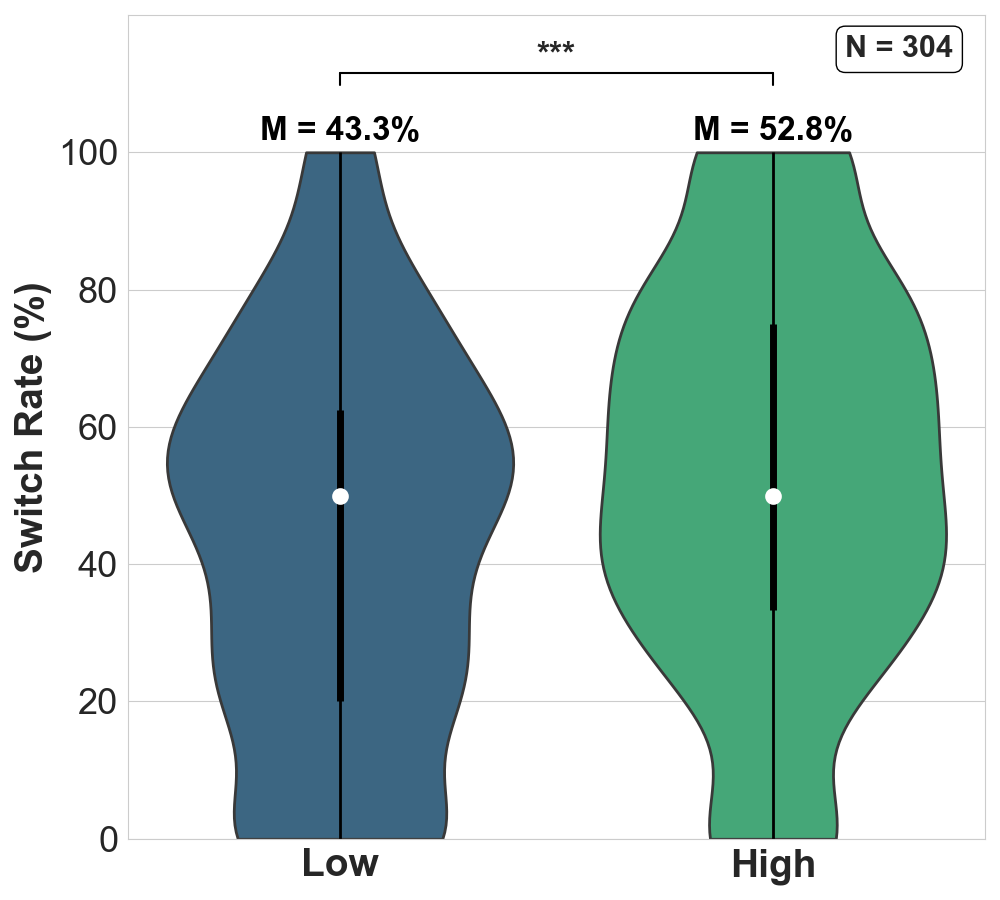}
        \caption{Detail Level}
        \label{fig:switch_rate_detail}
    \end{subfigure}
        \hfill
    \begin{subfigure}{0.32\textwidth}
        \includegraphics[width=\linewidth]{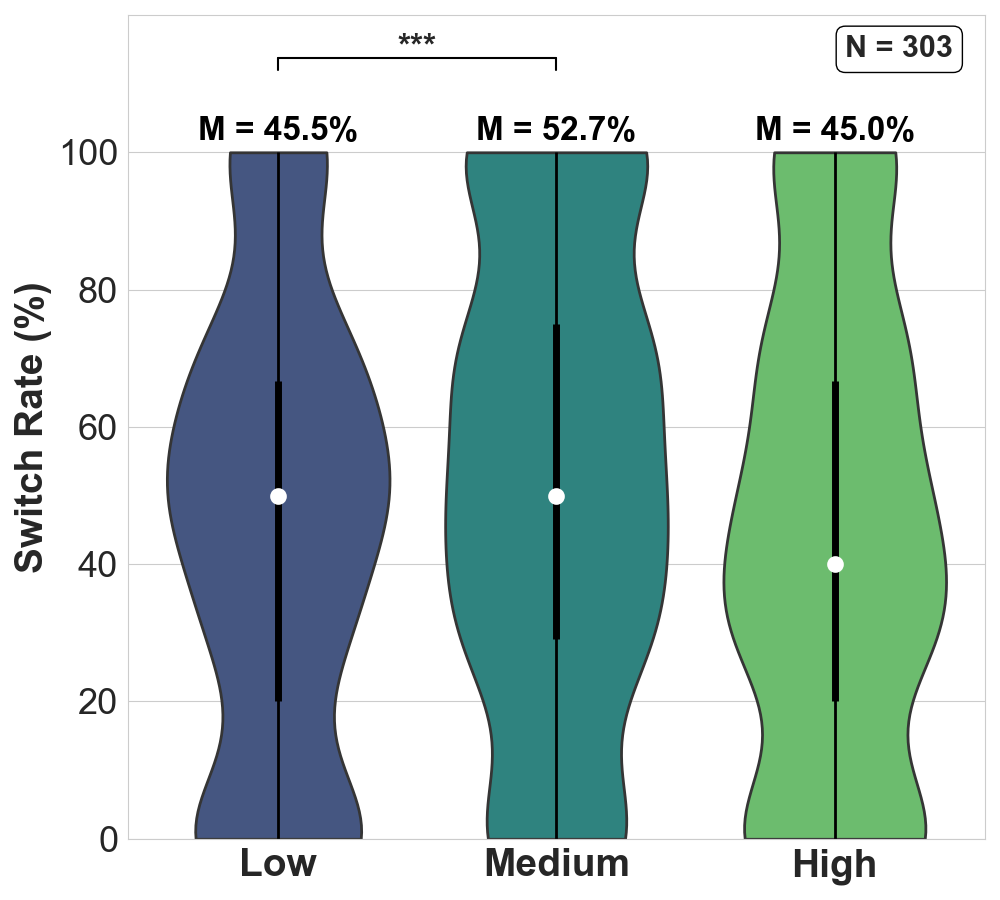}
        \caption{Confidence Level}
        \label{fig:switch_rate_confidence}
    \end{subfigure}
    \caption{Belief switch rate across different conditions. \zekun{The switch rate indicates the percentage of trials where participants switched their stance when presented with an opposing AI  response. This rate is lower in the Opinion Evaluation task, and significantly affected by the level of detail and confidence level of the AI response. Numbers above the distribution represent the mean value.  }}
    \label{fig:switch_rate}
\end{figure}

\zekun{To better illustrate this effect, we computed each participant’s \textit{switch rate} as the percentage of trials where participants switched their stance when presented with an opposing view of the AI. This switch rate corresponds to our definition of belief flip (as indicated in \autoref{tab:belief-change-types}) and allows us to examine the switch in trials where it was realistically expected. We visualize the distribution of participant-level switch rate across task types (\autoref{fig:switch_rate_task}), detail level (\autoref{fig:switch_rate_detail}), and confidence level (\autoref{fig:switch_rate_confidence}). The number above each distribution represents the mean  switch rate in each condition. 
}

In addition, participants’ initial belief strength ($\beta = -.89$, $SE = .06$, $p < .001$) and their perceived agreement with the AI ($\beta = -7.80$, $SE = .74$, $p < .001$) had a significant effect on belief switch. The more confident participants were in their original belief,  the less likely they were to change their stance and participants who perceived their stance to align with that of the AI were less likely to switch their belief compared to when they disagreed with it. 

In summary, we find that users were more likely to change their belief stance when the AI response provided higher information detail and was framed with a neutral, moderately confident tone. These two features independently increased belief switch rates, although their combined effect was weaker than expected. In contrast, highly confident responses were not more persuasive than low-confidence ones.


\subsubsection{Belief Shift}
\label{subsec:belief_shift}

\begin{figure}[htbp]
    \centering
    \begin{subfigure}{0.32\textwidth}
        \includegraphics[width=\linewidth]{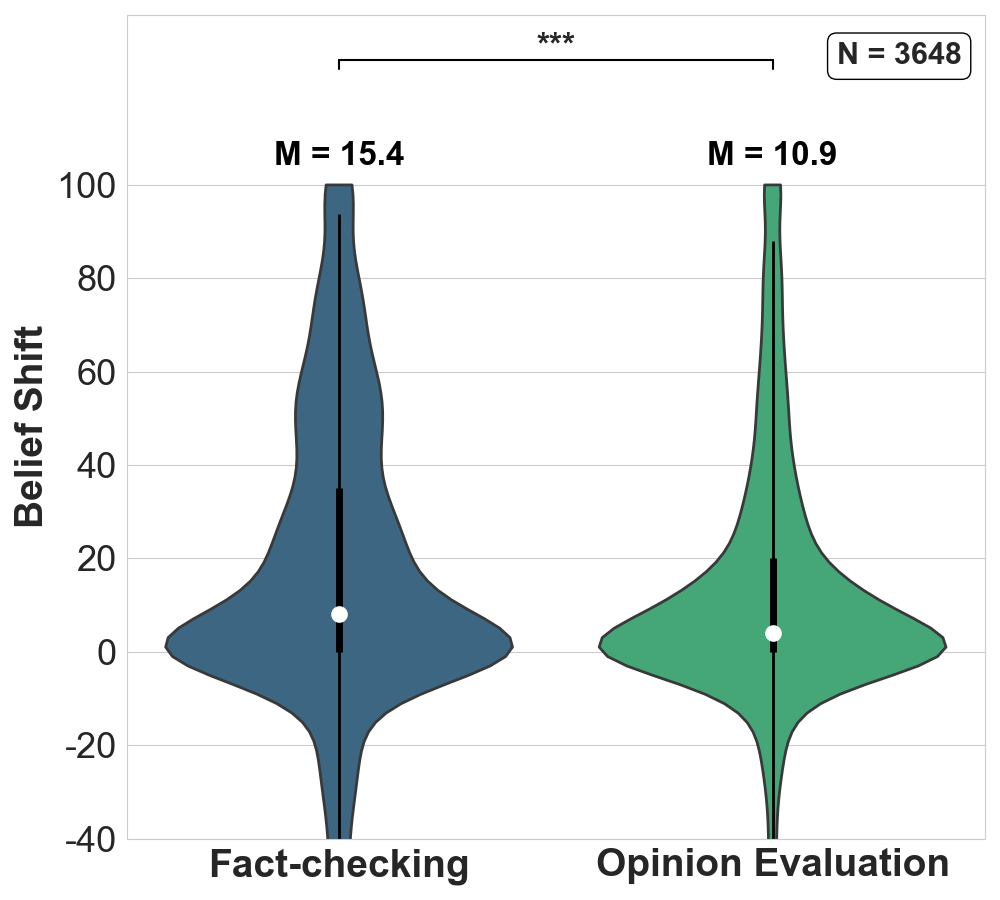}
        \caption{Task Type}
        \label{fig:shift_task}
    \end{subfigure}
    \hfill
    \begin{subfigure}{0.32\textwidth}
        \includegraphics[width=\linewidth]{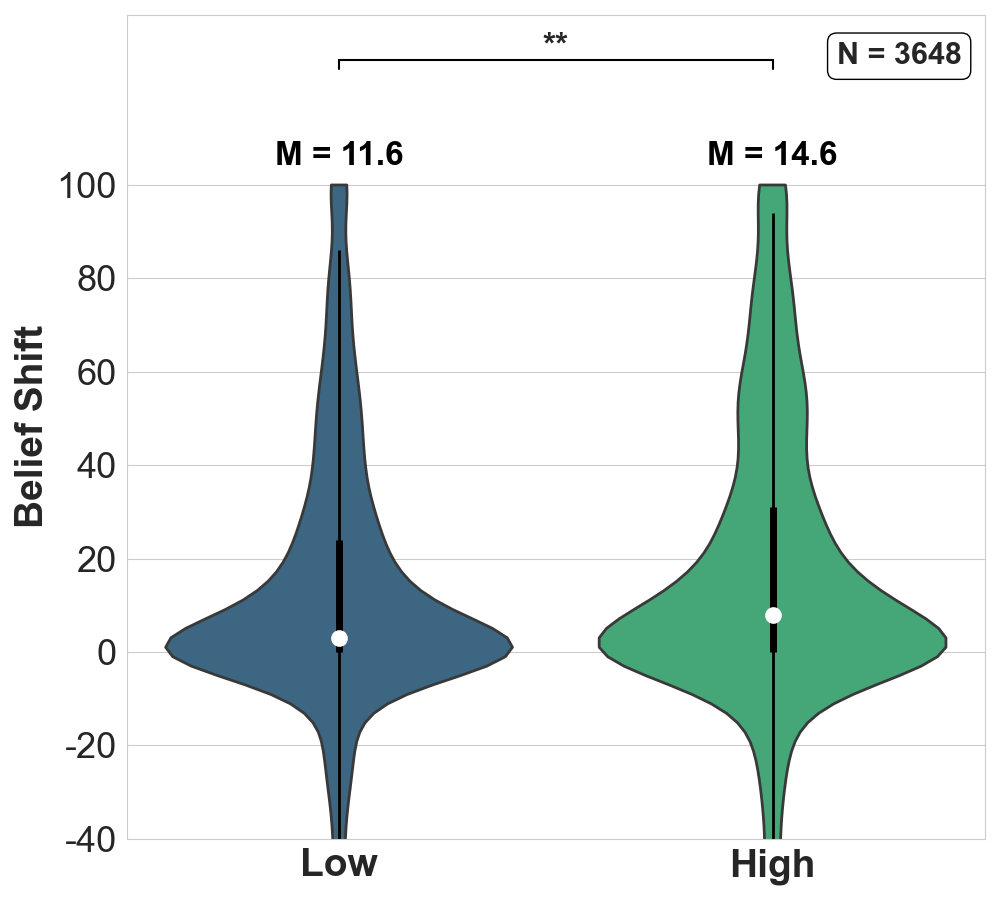}
        \caption{Detail Level}
        \label{fig:shift_detail}
    \end{subfigure}
        \hfill
    \begin{subfigure}{0.32\textwidth}
        \includegraphics[width=\linewidth]{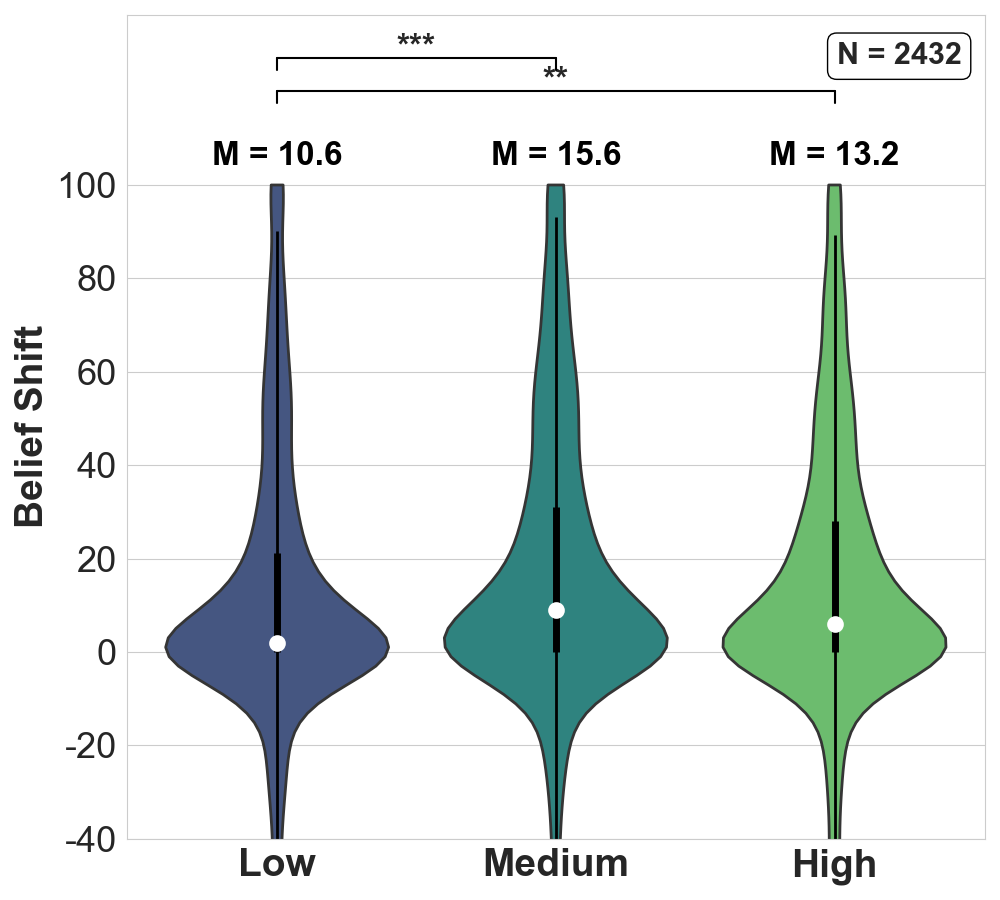}
        \caption{Confidence Level}
        \label{fig:shift_confidence}
    \end{subfigure}
    \caption{\zekun{The continuous belief shift metric captures the magnitude of change in the participants' certainty about a claim or opinion after seeing the AI's response, with a positive value indicating a shift towards the AI's stance. Our study shows a substantial weakening or reinforcement of a participant's original stance towards the AI's stance, which is affected by the task type, the level of detail, and the confidence level of the AI's response. Numbers above the distribution represent the mean value. }}
    \label{fig:shift}
\end{figure}

    To understand how participants adjusted the strength of their beliefs in response to AI suggestions, we analyzed the \textit{belief shift} measure, which captures the extent of movement toward or away from the AI’s stance, as measured by the change in belief strength. We calculated this shift at the trial level.
    and visualized the distribution across task type (\autoref{fig:shift_task}), detail level (\autoref{fig:shift_detail}), and confidence level (\autoref{fig:shift_confidence}). 
    


\zekun{As indicated in the \autoref{fig:shift_task}, belief shift was significantly influenced by task context. Participants showed a stronger movement towards AI’s stance in the fact-checking task ($M = 15.4$) than in the opinion evaluation task ($M = 10.9$), with a main effect of the task ($\beta = -6.57$, $SE = 1.29$, $p < .001$). This indicates that participants were more likely to align their beliefs with the AI’s stance when the claim could be objectively verified. }

\zekun{For the detail level,  responses with higher information detail led to a greater shift towards the AI-suggested stance ($M = 11.6$) compared to low detail responses ($M = 11.6$), as indicated in \autoref{fig:shift_detail}. We found a significant main effect of the detail level ($\beta = 3.72$, $SE = 1.16$, $p = .001$). The confidence level had an influence as well (\autoref{fig:shift_confidence}): compared to low-confidence responses ($M = 10.6$), both medium-confidence ($M = 15.6$; $\beta = 5.84$, $SE = 1.16$, $p < .001$) and high-confidence messages ($M = 13.2$; $\beta = 3.42$, $SE = 1.16$, $p = .003$) increased belief shift, with the strongest effect observed for medium confidence. } We did not observe a significant interaction effect on belief shift between detail and confidence level.

\zekun{\autoref{fig:shift} shows the distribution of belief shifts across conditions. 
Note that the belief shift metric can theoretically reach -100, but values below -25 (indicating participants switched from agreeing with the AI to opposing it) were rare in our study. We therefore truncated the negative scale for clearer visualization.
In most trials, the change in belief strength ranges from ca. 0 to +25. Although this number may appear small, it can indicate a substantial weakening or strengthening of participants’ beliefs relative to the AI’s stance, given that belief strength was measured on a 0–50 scale. 
In addition, most distributions show a second minor mode around the 50 point mark, which represents trials where participants switched their original stance towards the AI's stance, resulting in an absolute change in belief strength of more than 50 points. }

In addition to task and message framing, several other factors significantly influenced belief shift. Belief shift towards the AI's stance was lower when participants held stronger initial beliefs ($\beta = -3.11$, $SE = .39$, $p < .001$), perceived agreement with the AI's stance compared to disagreeing with the AI ($\beta = -25.28$, $SE = .71$, $p < .001$), or encountered the AI message earlier in the session ($\beta = .15$, $SE = .05$, $p = .001$). 

\zekun{To better understand how people shift their beliefs, we further examined how the influence of AI responses on belief change depends on the strength of participants’ initial beliefs. To simplify, we dichotomized initial belief strength by measuring the distance of a participant’s initial rating from the scale midpoint (50). Distances greater than 25 were classified as high initial belief strength, and distances of 25 or less as low initial belief strength.}

\begin{figure}[htbp]
    \centering
    \begin{subfigure}{0.32\textwidth}
        \includegraphics[width=\linewidth]{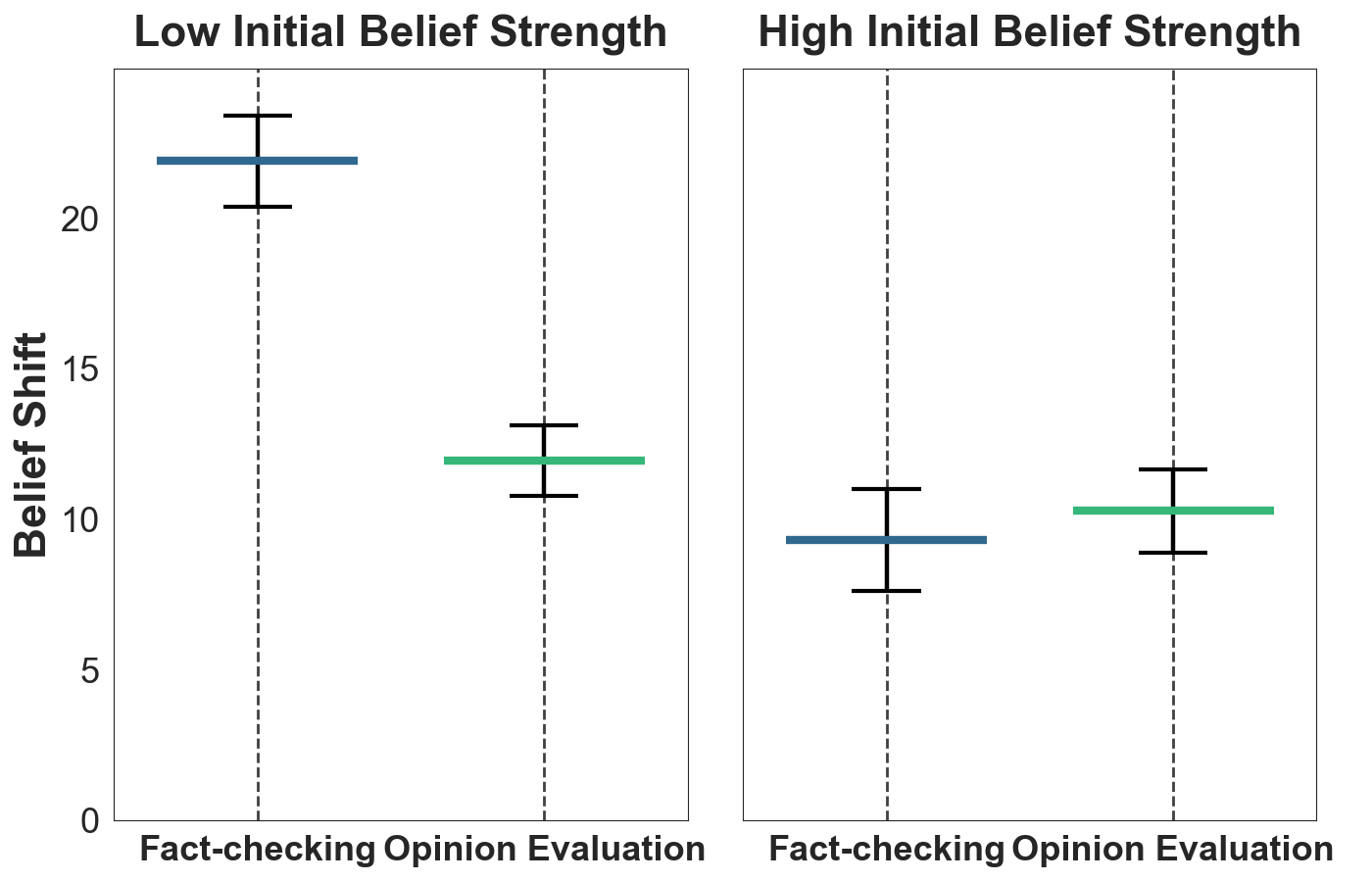}
        \caption{Task Type}
        \label{fig:initial_certainty_task}
    \end{subfigure}
    \hfill
    \begin{subfigure}{0.32\textwidth}
        \includegraphics[width=\linewidth]{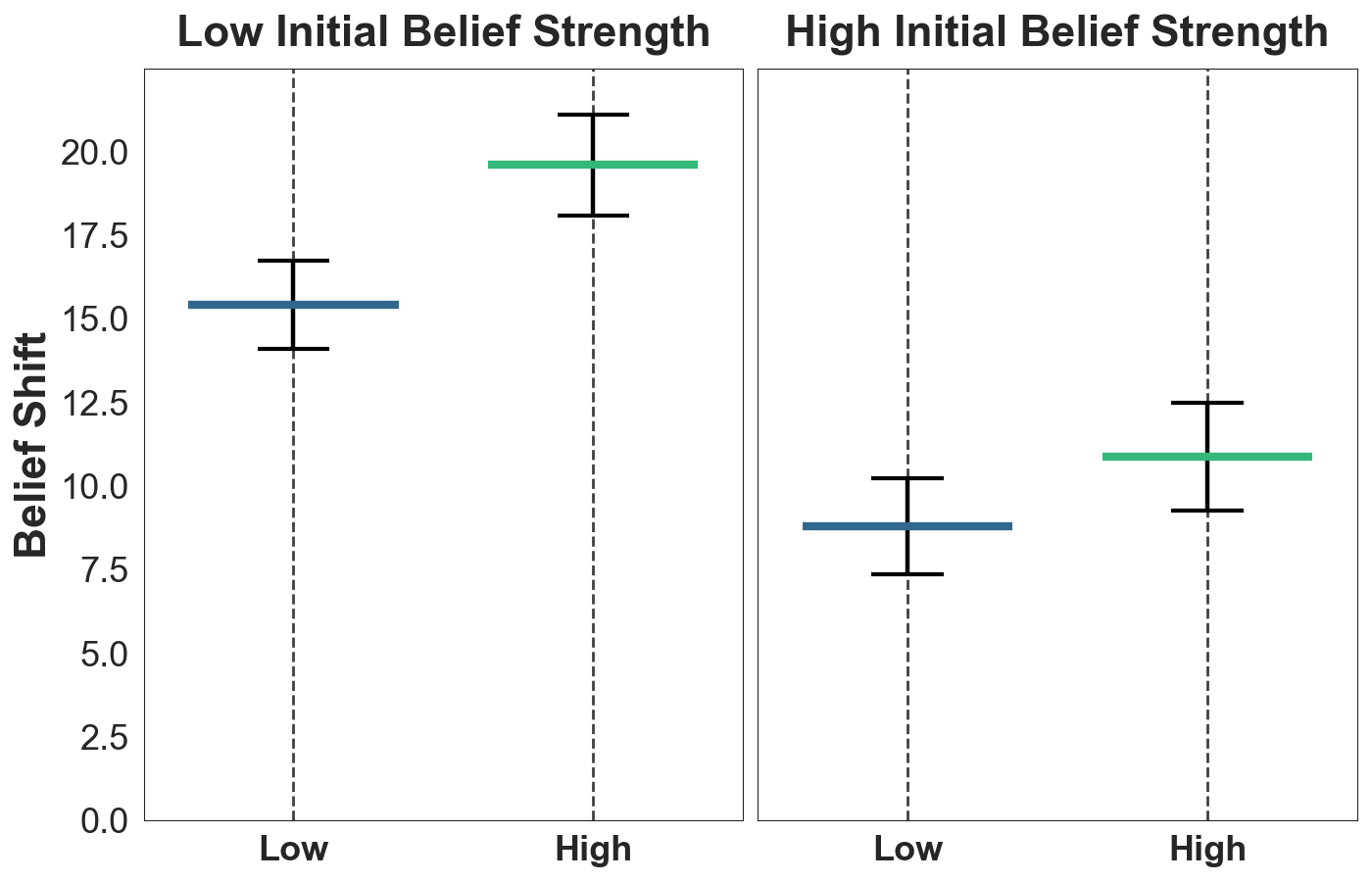}
        \caption{Detail Level}
        \label{fig:initial_certainty_detail}
    \end{subfigure}
        \hfill
    \begin{subfigure}{0.32\textwidth}
        \includegraphics[width=\linewidth]{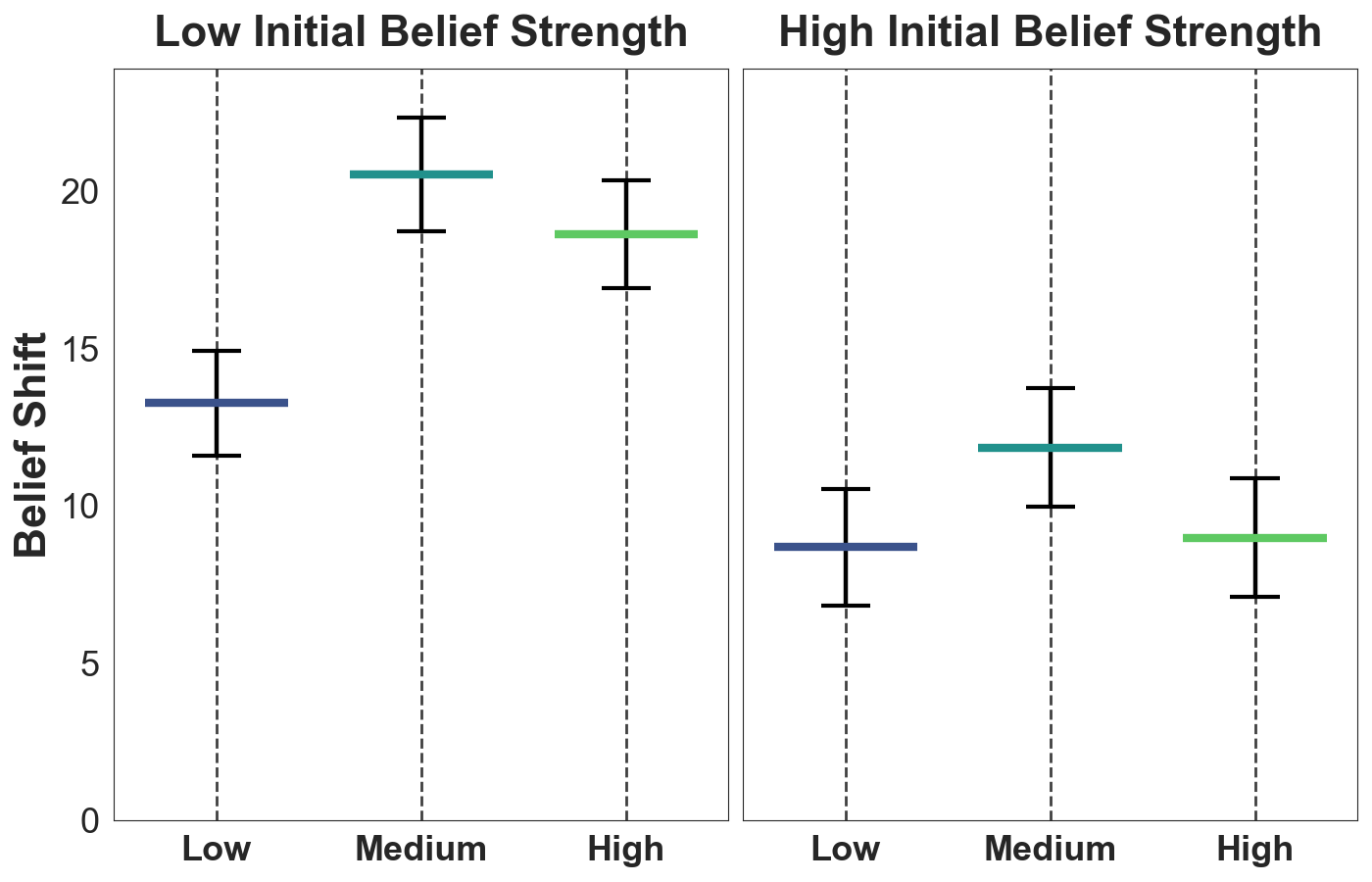}
        \caption{Confidence Level}
        \label{fig:initial_certainty_confidence}
    \end{subfigure}
    \caption{Belief Shift across different conditions, split by initial belief strength}
    \label{fig:initial_certainty}
\end{figure}

\zekun{As shown in \autoref{fig:initial_certainty}, belief shifts were consistently larger when participants held weaker initial beliefs. For example, in the fact-checking task, users with low initial belief strength (left of \autoref{fig:initial_certainty_task}) showed substantially greater shifts than those with high initial belief strength  (right of \autoref{fig:initial_certainty_task}). A similar pattern can be observed for both detail and confidence conditions: differences between conditions were much more pronounced when initial beliefs were weak (left of \autoref{fig:initial_certainty_detail} and \autoref{fig:initial_certainty_confidence}). In addition, the effect of confidence on belief shift was amplified under low initial belief strength, with medium-confidence responses producing the strongest changes.}

In summary, AI responses that included high levels of detail and were expressed with either neutral or high confidence were more effective in shifting users’ beliefs toward the AI’s stance. Among these, messages with neutral confidence showed the strongest effect, suggesting that a balanced tone may enhance persuasive impact without triggering resistance. 

\subsubsection{Perceived AI Influence}
\label{subsec:perceived_influence}

\begin{figure}[htbp]
    \centering
    \begin{subfigure}{0.32\textwidth}
        \includegraphics[width=\linewidth]{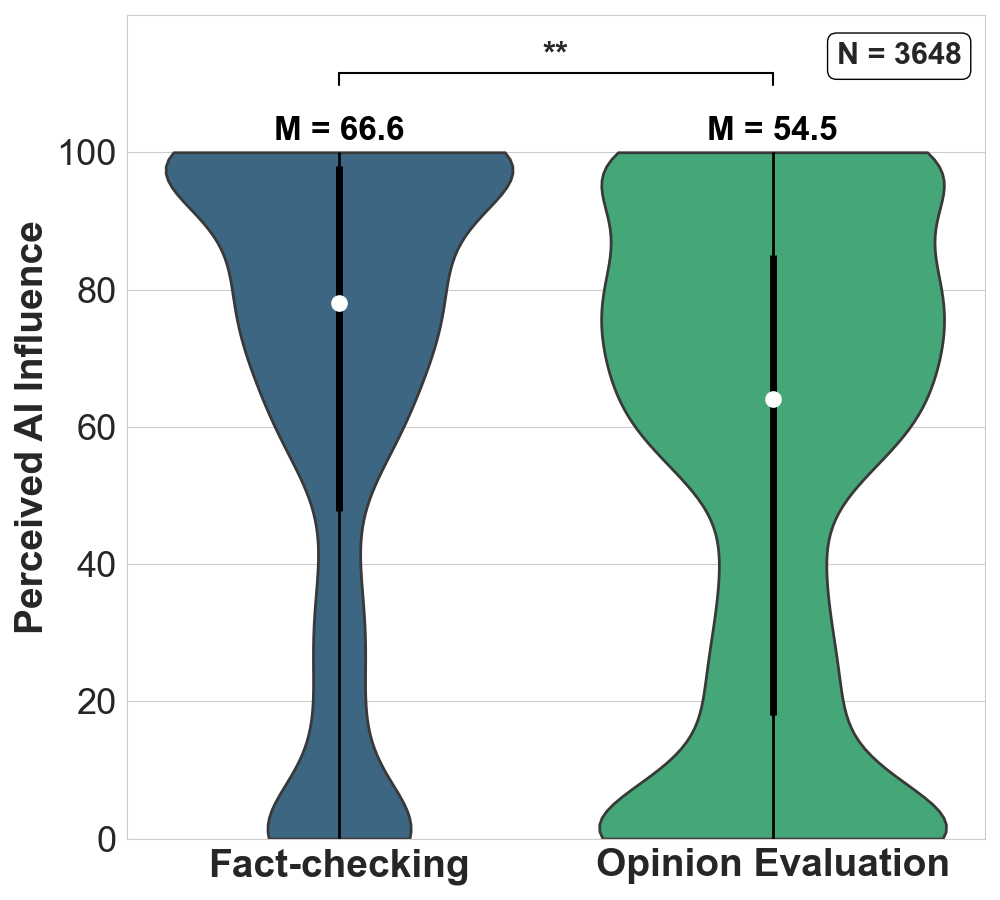}
        \caption{Task Type}
        \label{fig:influence_task}
    \end{subfigure}
    \hfill
    \begin{subfigure}{0.32\textwidth}
        \includegraphics[width=\linewidth]{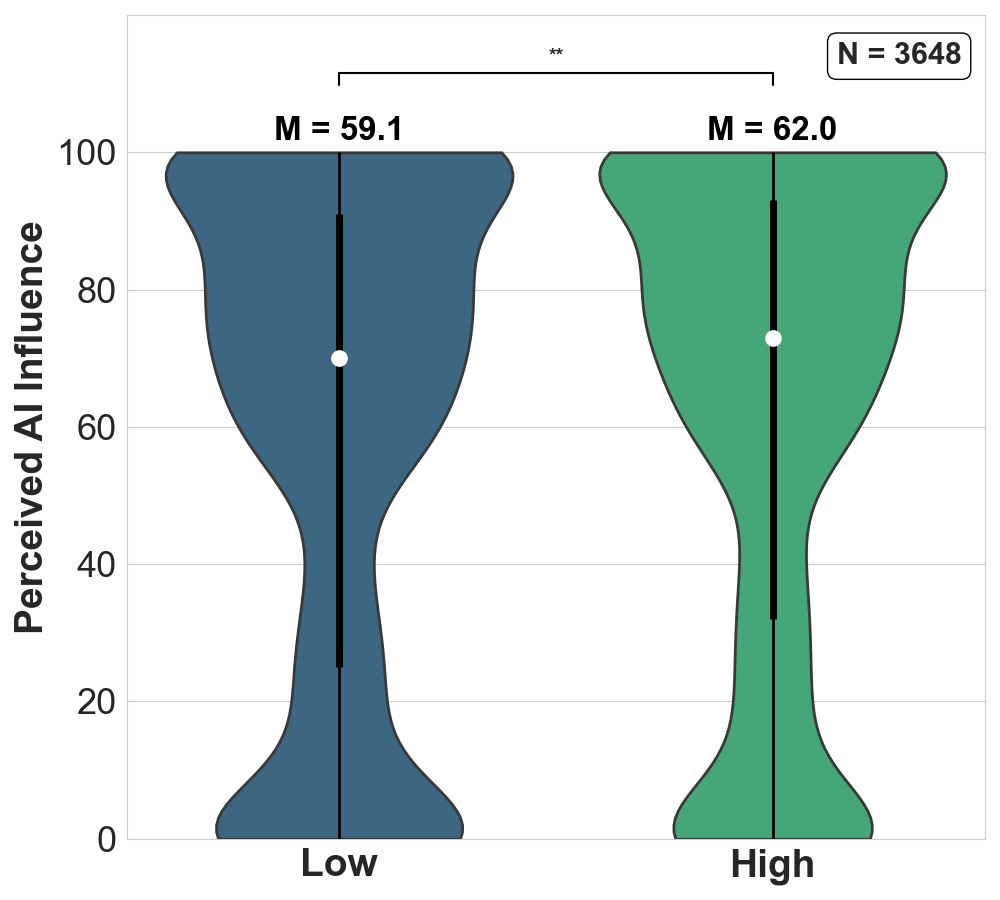}
        \caption{Detail Level}
        \label{fig:influence_detail}
    \end{subfigure}
        \hfill
    \begin{subfigure}{0.32\textwidth}
        \includegraphics[width=\linewidth]{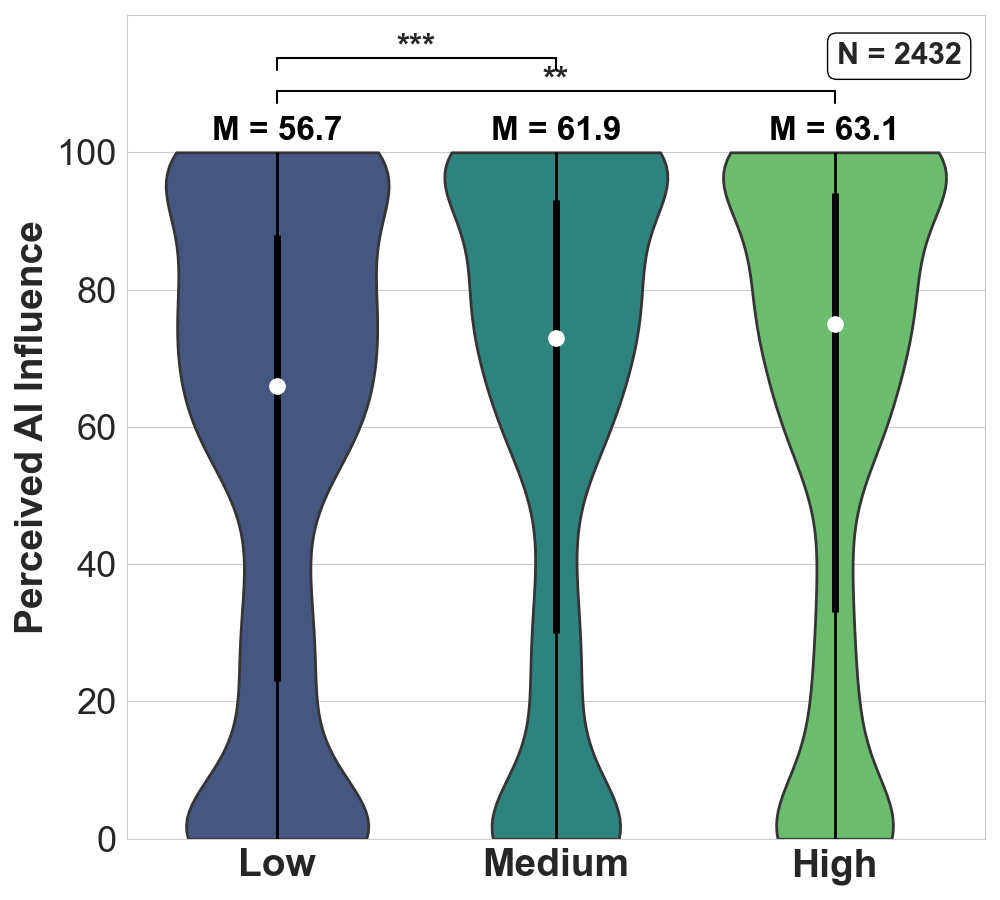}
        \caption{Confidence Level}
        \label{fig:influence_confidence}
    \end{subfigure}
    \caption{Perceived AI Influence in Different Condition}
    \label{fig:belief_influence}
\end{figure}

We also examined participants’ subjective perception of how much the AI response influenced their final decision. We analyzed perceived AI influence across task type (\autoref{fig:influence_task}), information detail level (\autoref{fig:influence_detail}), and confidence level (\autoref{fig:influence_confidence}). 

Participants reported significantly greater AI influence in the fact-checking task compared to the opinion evaluation task ($M=66.59$ vs $54.53$, $\beta = -9.94$, $SE = 0.80$, $p < .001$). Higher detail in the AI response also led to greater perceived influence ($M=62.01$ vs $59.11$, $\beta = 2.21$, $SE = 1.11$, $p = 0.05$). Confidence level also influenced perceived AI impact: compared to low-confidence messages ($M=56.65$), AI responses framed with medium confidence ($M=63.11$, $\beta = 3.50$, $SE = 1.11$, $p = .001$) or high confidence ($M=61.93$, $\beta = 6.02$, $SE = 1.11$, $p < .001$) were associated with significantly higher perceived influence.

In addition to task type and message framing, other factors also significantly influenced participants’ perceived AI influence. Participants reported less influence from the AI on trials where they held stronger initial beliefs ($\beta = -10.61$, $SE = .40$, $p < .001$), perceived disagreement with the AI stance ($\beta = -4.13$, $SE = .68$, $p < .001$), or encountered the AI message earlier in the session ($\beta = .11$, $SE = .05$, $p = .012$). 

\subsubsection{Summary}
\label{subsec:summary}

Across all three belief change measures —belief switch, belief shift, and perceived AI influence—we consistently found that AI responses with higher information detail and moderate confidence were more persuasive. High detail increased the likelihood of belief switch, promoted stronger alignment with the AI’s stance, and enhanced perceived influence. Similarly, messages framed with medium confidence were more effective than low or high confidence in encouraging belief change across all measures. Interestingly, while high confidence increased belief shift and perceived influence, it did not significantly affect belief switch, suggesting that assertive tone may enhance conviction but not necessarily prompt users to reverse their stance. These results highlight the nuanced impact of AI communication style on users’ belief formation and adjustment.

\subsection{Exploratory Modeling: Misperception and Belief Deviation}
\label{sec:deviation}

For the final part of the analysis, we explored potential explanations for two phenomena: participants’ misperception of the AI-generated response and the more perplexing cases of belief deviation—where participants updated their belief in a direction opposite to the AI’s stance.

\begin{table}[htbp]
\centering
\caption{Misperception Cases by Detail and Confidence Level (Counts and Percentages)}
\label{tab:misperception_margin}
\footnotesize
\begin{tabular}{lccc|c}
\toprule
\multirow{2}{*}{\textbf{Detail Level}} & \multicolumn{3}{c|}{\textbf{Confidence Level}} & \multirow{2}{*}{\textbf{Total}} \\
\cmidrule(lr){2-4}
& \textbf{Low} & \textbf{Medium} & \textbf{High} & \\
\midrule
Low  & 121 (17.0\%) & 102 (14.3\%) & 127 (17.8\%) & 350 (49.2\%) \\
High & 125 (17.6\%) & 116 (16.3\%) & 121 (17.0\%) & 362 (50.8\%) \\
\midrule
\textbf{Total} & 246 (34.6\%) & 218 (30.6\%) & 248 (34.8\%) &  \\
\bottomrule
\end{tabular}
\end{table}

\begin{table}[htbp]
\centering
\caption{Deviation Cases by Detail and Confidence Level (Counts and Percentages)}
\label{tab:deviation_margin}
\footnotesize
\begin{tabular}{lccc|c}
\toprule
\multirow{2}{*}{\textbf{Detail Level}} & \multicolumn{3}{c|}{\textbf{Confidence Level}} & \multirow{2}{*}{\textbf{Total}} \\
\cmidrule(lr){2-4}
& \textbf{Low} & \textbf{Medium} & \textbf{High} & \\
\midrule
Low  & 279 (19.8\%) & 227 (16.1\%) & 201 (14.3\%) & 707 (50.2\%) \\
High & 274 (19.5\%) & 202 (14.4\%) & 224 (15.9\%) & 700 (49.8\%) \\
\midrule
\textbf{Total} & 553 (39.3\%) & 429 (30.5\%) & 425 (30.2\%) &  \\
\bottomrule
\end{tabular}
\end{table}

The overall misperception rate across all participants was \textit{9.8\%} with a total of 712 cases. To investigate whether this was related to the manipulated features of the AI responses, we first visualized the distribution of misperception cases across the six experimental conditions. As shown in \autoref{tab:misperception_margin}, misperception appears fairly evenly distributed across all combinations of detail and confidence levels, with no obvious concentration in specific conditions. To further examine this, we ran a mixed-effects logistic regression, using the same model structure described in \autoref{sec:belief_change_effects_test}, but replacing the dependent variable BM with a binary indicator of misperception (1 = misperceived, 0 = not). We excluded the independent variable \(\mathit{Perceived\_AI\_Agreement}\), as it directly overlaps with the misperception definition, but retained all other predictors. The results showed that initial belief strength was the most significant predictor of misperception ($\beta = -0.14$, $SE = .05$, $p = .007$), suggesting that the less certain participants were about the claim initially, the more likely they were to misinterpret AI’s stance. A smaller but still significant effect of task type was also found ($M=9.7\%$ vs $9.8\%$,$\beta = 0.41$, $SE = .19$, $p = .03$), indicating that misperception occurred more frequently in the opinion evaluation task than in the fact-checking task. No significant main effects of detail level, confidence level, or their interaction were observed.

Turning to belief deviation, we found that \textit{19.3\%} of all trials resulted in belief updates that went against the actual AI stance, as illustrated in \autoref{fig:ai_influence}. Unlike misperception, these deviation cases were less evenly distributed, as indicated in \autoref{tab:deviation_margin}. For instance, low-confidence conditions—whether paired with high detail (19.5\%) or low detail (19.8\%)—showed noticeably higher deviation rates (39.3\%).

To explore the relationship between misperception and belief deviation, we performed another mixed-effects logistic regression, this time using belief deviation as a binary dependent variable and including misperception as an additional predictor in the model used in \autoref{sec:belief_change_effects_test}. We found that the confidence level in LLMs' generated responses did have a significant effect on turning participants' against the  AI stance: compared to low-confidence messages ($M = 22.7\%$), AI responses expressed with medium confidence ($M =17.6\%$) were more likely to lead to belief deviation ($\beta = .76$, $SE = .18$, $p < .001$). This effect was not observed for highly confident messages ($M = 45.1\%$), suggesting that overconfidence may reduce persuasive impact, which is consistent with previous findings \cite{xu2025confronting}.
The results revealed significant effects of both misperception ($M=0.75$ vs $0.13$,$\beta = 4.15$, $SE = .13$, $p < .001$) and initial belief strength ($\beta = -0.15$, $SE = .05$, $p < .001$), indicating that participants who misunderstood the AI’s stance and those who were initially less certain were significantly more likely to update their belief in a direction opposite to the AI suggestion.

\section{Discussion}
\label{sec:discussion}
In this part, we first highlight our key findings related to AI message characteristics and user factors that shape belief change, then discuss broader implications for evaluation metrics, system design, and ethical concerns. We conclude with limitations of our study.

\subsection{Key Findings}
\textbf{Detailed and confident AI responses drive belief change, but overconfidence expression discourages belief switch.} Specifically, we observed consistent effects of information detail in AI responses across all three belief change measures: belief switch, belief shift, and perceived AI influence. When the AI message included more factual content—such as citations, key facts, and statistics—participants were more likely to reverse their initial stance, shift their beliefs toward the AI’s position, and report stronger influence from the AI on their final judgment. These findings support our hypothesis regarding the role of detail level (H2) and align with prior work showing that well-supported explanations improve user receptiveness to AI-generated content~\cite{kim2025fostering}.

For confidence level, we found that responses expressed with medium or high confidence were more effective at shifting participants’ beliefs compared to low-confidence responses, partially supporting our hypothesis that higher confidence increases belief change (H3). However, only medium-confidence messages significantly increased belief switch rates. Thus, while high-confidence responses shifted beliefs, this did not lead to full reversals of stances. This means that highly assertive language can contribute to reinforcing or weakening beliefs to align more strongly with the AI' stance but is less likely to fully overturn a decision outcome. 
\zekun{Nevertheless, participants \emph{perceived} the AI's influence on their belief to be similar when expressed with high or medium confidence. Only in the low-confidence tone was the AI influence perceived as less by the participants. }
\zekun{Importantly, our continuous belief shift metric allowed us to uncover these subtle but important effects of how different confidence expressions shape users' beliefs, even if they do not result in different decisions.}


\textbf{Subjective and context factors, including the objectivity of the task and the initial belief strength, drive belief change to a large extent.} Across the two task scenarios in our study, participants were more likely to switch their belief stance, shift closer to the AI’s suggested stance, and report stronger AI influence in the fact-checking task compared to the opinion evaluation task, which confirms our hypothesis that users may be more receptive to AI suggestions when claims can be objectively verified (H1). \zekun{Belief shifts were also stronger when participants were less certain about their initial belief. This could indicate that people with lower conviction might be more susceptible to AI influence and again highlight the importance of understanding belief shift on a continuous scale rather than only looking at binary outcome measures.}

\textbf{Uncertainty markers in AI response and misperception of AI's stance can drive deviations from AI's stance.} Our exploratory analysis highlights that belief change is not solely determined by the content of AI responses, but also by how users perceive the AI's stance. Misperception—when users misidentify whether the AI agrees or disagrees with their initial belief stance—was observed in nearly 10\% of trials. This mismatch between actual AI suggested stance and the users' initial belief stance can further have a significant effect on belief deviation—cases in which users updated their post-AI belief in a direction that contradicted the AI's stance. In addition to misperception, belief deviation was more common following AI responses expressed with low confidence. Compared to responses with moderate or high confidence, low-confidence responses increased the likelihood that users would shift away from the AI’s stance, which suggests that expressions of uncertainty may erode users’ trust in the AI, causing them to dismiss or reject the suggested stance passed in the message. \zekun{Similar to previous findings, this shift might not result in a reversal of stance and can thus only be analyzed using a continuous belief shift metric as the one introduced in our analysis framework}.

\subsection{Broader Implication}
\subsubsection{Rethinking Evaluation Metrics for AI Influence on Human Behavior and Belief} 
Through our study, we found that the influence of AI on user belief extends well beyond binary outcome measures. Evaluating this influence should not be limited to one-dimensional comparisons—such as whether the user accepted the AI's answer—but must also account for both the direction and magnitude of belief change. Without more nuanced measures that capture these variations, we risk misinterpreting how users engage with AI-generated content and form their post-AI beliefs. Even when the final decision appears unchanged, the underlying belief dynamics can differ substantially: users may strengthen or weaken their original belief in line with the AI’s suggestion without shifting their stance, or even change their belief against the AI's stance (deviations). 

In this study, we introduced the \textit{belief shift} measure as a complementary approach to the commonly used binary measure. By quantifying the extent to which users move towards or away from AI’s stance, the shift measure enables a deeper and more comprehensive analysis of human-AI interaction—particularly in open-ended or ambiguous tasks. For example, in political or social discourse, an individual may not change their stance on a controversial issue, but their beliefs might shift closer to the opposing side, reflecting increased openness or reduced polarization. Similarly, in educational contexts, a student might retain the same answer to a question but update their confidence in that answer—indicating meaningful learning that traditional correctness metrics would miss. 

As an alternative to the continuous belief shift measure, we also proposed a categorization of belief changes that not only considers belief flips but also belief reinforcements and belief weakening to capture more subtle belief changes towards the AI's stance, as well as categorizing cases where participants change their beliefs to deviate from the AI's stance. This seemingly counterintuitive behavior could be partially explained by low-confidence responses leading to misperceptions and hints towards mistrust in AI responses. Thus, we recommend that future empirical research incorporate such measures to more accurately assess the nature and depth of AI’s influence on human belief.

\subsubsection{Designing User-Centered, Belief-Aware LLM Systems}
The answer to the question: \textit{“How do the information detail and confidence level of the AI-generated responses influence users’ belief stance and strength during interaction?”}  carry important implications for the design of future LLM-powered systems, particularly those that support information consumption and decision-making.

\zekun{Our findings highlight important ethical challenges: even when unintended, AI responses can nudge users’ convictions in ways that may affect autonomy. This makes it essential to approach the design of belief-aware systems with caution. In particular, seemingly small choices, such as how much detail an AI provides or how confidently it phrases its answer, can subtly influence both the stance and the strength of belief. One practical implication is that confidence expressions in LLM outputs should be calibrated to more accurately reflect the model’s underlying certainty, reducing the risk of overstating or understating reliability. There have been first attempts toward this goal, such as evaluating the reliability of uncertainty estimates across large sets of LLMs \cite{tao2025revisiting}, modeling certainty phrases as probability distributions to better align expression with accuracy \cite{wang2024calibrating}, or providing explanations that show where evidence supports or contradicts a claim \cite{sun2025explaining}. Our results complement these directions by demonstrating why calibration matters: uncalibrated confidence cues not only influence trust, but can also shift users’ convictions. A key challenge remains that estimating the model’s true uncertainty for a specific output is inherently difficult, which complicates attempts to tune the expression precisely. Nonetheless, connecting technical calibration methods with empirical insights on belief change can help future systems align their tone more closely with their reliability and avoid overreliance.}

\subsubsection{Ethical Concern for Malicious Uses of AI Influence on Human Belief} 

On the negative side, the very ability of AI-generated responses to influence user beliefs opens the door for malicious actors to exploit this power for harmful purposes. Persuasive language styles can be weaponized to spread misinformation, subtly reinforce conspiracy theories, or manipulate individuals into making detrimental decisions—such as promoting risky financial behavior or discouraging necessary medical interventions. A very recent study by Li et al. \cite{li2025text} demonstrates that LLM-generated analyses, even when based on the same decision recommendation, can be selectively presented to nudge human decisions by emphasizing different supporting aspects. Our findings further suggest that users can internalize the content of such AI-generated messages, forming new post-AI beliefs that could persist well beyond a single decision point. This highlights the risk that LLMs may not merely influence actions in the moment but can also reshape longer-term attitudes and convictions. Therefore, understanding and addressing the ethical implications of belief manipulation is critical. Future research should focus not only on enhancing the supportive capabilities of AI systems but also on developing safeguards to prevent the misuse of their persuasive power, ensuring that AI assists rather than exploits users in their belief formation processes. 

\subsection{Limitation}

Our study has several limitations. First, as a survey-based experiment, participants did not have the opportunity to actively interact with the LLM, such as posing their own questions or inquiries to their specific interests. Instead, they were presented with pre-generated AI responses and instructed not to consult external resources. This controlled setup, while necessary for isolating specific experimental manipulations, limits how well the findings reflect real-world AI use. 

Second, we did not examine the long-term effects of AI responses on belief change. It remains unclear whether the observed shifts in belief persist over time or whether repeated exposure to consistent AI responses could gradually influence users’ beliefs, particularly in cases where participants initially held strong convictions.


\zekun{Third, belief change in our study may have been shaped by two factors we did not directly control. On the one hand, participants likely differed in their tendency to trust or rely on AI support, which we did not measure. On the other hand, the two tasks exposed participants to different patterns of AI accuracy: in fact-checking the AI was always correct (though not disclosed), while in opinion evaluation the AI responses were evenly split between support and opposition. These individual and task-level differences could have influenced not only participants’ trust in the AI but also the extent of their belief change, especially if early trials created strong first impressions.}

\zekun{Finally, our study isolates detail and confidence features but does not address other forms of persuasive framing such as emotional tone, personalization, or deceptive explanations. Future work should examine these features under an explicit ethical lens, as they may present greater risks of manipulation.}

\section{Conclusion}
\label{sec:conclusion}
In this paper, we present empirical findings from a pre-registered online experiment (N =304) that investigates how two key features of AI-generated responses—information detail and confidence level—influence user belief change. To capture how users’ belief stance and belief strength evolve before and after engaging with AI, we introduce a comprehensive analytical framework that classifies different belief change outcomes and two complementary belief change measures: \textit{belief switch}, which captures whether users reversed their belief stance, and \textit{belief shift}, which quantifies the extent to which users moved toward or away from AI’s suggestion.

Our results show that increasing the level of detail or adopting a moderately confident tone in AI responses can effectively trigger belief switches, encourage belief shifts toward the AI’s stance, and increase perceived AI influence on final decision. However, overly confident language with definitive statements may discourage users from changing their belief stance, while low-confidence responses may prompt users to shift away from the AI’s suggestion due to perceived unreliability. Beyond the response features themselves, other factors—including task context, initial belief strength, and perceived agreement with the AI—also played a decisive role in shaping user beliefs when interacting with AI. All these findings highlight the importance of designing LLM systems that can not only inform users, but also responsibly guide belief formation in ways that are context-aware and aligned with human understanding.



\bibliographystyle{ACM-Reference-Format}
\bibliography{reference}

@article{ferrucci2010building,
  title={Building Watson: An overview of the DeepQA project},
  author={Ferrucci, David and Brown, Eric and Chu-Carroll, Jennifer and Fan, James and Gondek, David and Kalyanpur, Aditya A and Lally, Adam and Murdock, J William and Nyberg, Eric and Prager, John and others},
  journal={AI magazine},
  volume={31},
  number={3},
  pages={59--79},
  year={2010}
}

@book{croft2010search,
  title={Search engines: Information retrieval in practice},
  author={Croft, W Bruce and Metzler, Donald and Strohman, Trevor and others},
  volume={520},
  year={2010},
  publisher={Addison-Wesley Reading}
}

@article{quelle2024perils,
  title={The perils and promises of fact-checking with large language models},
  author={Quelle, Dorian and Bovet, Alexandre},
  journal={Frontiers in Artificial Intelligence},
  volume={7},
  pages={1341697},
  year={2024},
  publisher={Frontiers Media SA}
}

@inproceedings{kim2025fostering,
author = {Kim, Sunnie S. Y. and Vaughan, Jennifer Wortman and Liao, Q. Vera and Lombrozo, Tania and Russakovsky, Olga},
title = {Fostering Appropriate Reliance on Large Language Models: The Role of Explanations, Sources, and Inconsistencies},
year = {2025},
isbn = {9798400713941},
publisher = {Association for Computing Machinery},
address = {New York, NY, USA},
url = {https://doi.org/10.1145/3706598.3714020},
doi = {10.1145/3706598.3714020},
booktitle = {Proceedings of the 2025 CHI Conference on Human Factors in Computing Systems},
articleno = {420},
numpages = {19},
location = {
},
series = {CHI '25}
}

@inproceedings{lai2023towards,
author = {Lai, Vivian and Chen, Chacha and Smith-Renner, Alison and Liao, Q. Vera and Tan, Chenhao},
title = {Towards a Science of Human-AI Decision Making: An Overview of Design Space in Empirical Human-Subject Studies},
year = {2023},
isbn = {9798400701924},
publisher = {Association for Computing Machinery},
address = {New York, NY, USA},
url = {https://doi.org/10.1145/3593013.3594087},
doi = {10.1145/3593013.3594087},
booktitle = {Proceedings of the 2023 ACM Conference on Fairness, Accountability, and Transparency},
pages = {1369–1385},
numpages = {17},
location = {Chicago, IL, USA},
series = {FAccT '23}
}

@article{goldstein24,
    author = {Goldstein, Josh A and Chao, Jason and Grossman, Shelby and Stamos, Alex and Tomz, Michael},
    title = {How persuasive is AI-generated propaganda?},
    journal = {PNAS Nexus},
    volume = {3},
    number = {2},
    pages = {pgae034},
    year = {2024},
    month = {02},
    issn = {2752-6542},
    doi = {10.1093/pnasnexus/pgae034},
    url = {https://doi.org/10.1093/pnasnexus/pgae034},
    eprint = {https://academic.oup.com/pnasnexus/article-pdf/3/2/pgae034/56712546/pgae034.pdf},
}

@article{bach2024systematic,
  title={A systematic literature review of user trust in AI-enabled systems: An HCI perspective},
  author={Bach, Tita Alissa and Khan, Amna and Hallock, Harry and Beltr{\~a}o, Gabriela and Sousa, Sonia},
  journal={International Journal of Human--Computer Interaction},
  volume={40},
  number={5},
  pages={1251--1266},
  year={2024},
  publisher={Taylor \& Francis}
}

@article{lee2023understanding,
  title={Understanding the effect of counterfactual explanations on trust and reliance on ai for human-ai collaborative clinical decision making},
  author={Lee, Min Hun and Chew, Chong Jun},
  journal={Proceedings of the ACM on Human-Computer Interaction},
  volume={7},
  number={CSCW2},
  pages={1--22},
  year={2023},
  publisher={ACM New York, NY, USA}
}

@article{jung2024quantitative,
  title={Quantitative Insights into Language Model Usage and Trust in Academia: An Empirical Study},
  author={Jung, Minseok and Zhang, Aurora and Lee, Junho and Liang, Paul Pu},
  journal={arXiv preprint arXiv:2409.09186},
  year={2024}
}

@article{karinshak2023working,
  title={Working with AI to persuade: Examining a large language model's ability to generate pro-vaccination messages},
  author={Karinshak, Elise and Liu, Sunny Xun and Park, Joon Sung and Hancock, Jeffrey T},
  journal={Proceedings of the ACM on Human-Computer Interaction},
  volume={7},
  number={CSCW1},
  pages={1--29},
  year={2023},
  publisher={ACM New York, NY, USA}
}

@inproceedings{xu2024role,
  title={On the role of large language models in crowdsourcing misinformation assessment},
  author={Xu, Jiechen and Han, Lei and Sadiq, Shazia and Demartini, Gianluca},
  booktitle={Proceedings of the International AAAI Conference on Web and Social Media},
  volume={18},
  pages={1674--1686},
  year={2024}
}

@article{lu2024does,
  title={Does more advice help? The effects of second opinions in AI-assisted decision making},
  author={Lu, Zhuoran and Wang, Dakuo and Yin, Ming},
  journal={Proceedings of the ACM on Human-Computer Interaction},
  volume={8},
  number={CSCW1},
  pages={1--31},
  year={2024},
  publisher={ACM New York, NY, USA}
}

@inproceedings{sharma2024generative,
  title={Generative echo chamber? effect of llm-powered search systems on diverse information seeking},
  author={Sharma, Nikhil and Liao, Q Vera and Xiao, Ziang},
  booktitle={Proceedings of the 2024 CHI Conference on Human Factors in Computing Systems},
  pages={1--17},
  year={2024}
}

@article{chen2023understanding,
  title={Understanding the role of human intuition on reliance in human-AI decision-making with explanations},
  author={Chen, Valerie and Liao, Q Vera and Wortman Vaughan, Jennifer and Bansal, Gagan},
  journal={Proceedings of the ACM on Human-computer Interaction},
  volume={7},
  number={CSCW2},
  pages={1--32},
  year={2023},
  publisher={ACM New York, NY, USA}
}

@article{park2021faviq,
  title={Faviq: Fact verification from information-seeking questions},
  author={Park, Jungsoo and Min, Sewon and Kang, Jaewoo and Zettlemoyer, Luke and Hajishirzi, Hannaneh},
  journal={arXiv preprint arXiv:2107.02153},
  year={2021}
}

@inproceedings{tan2016winning,
  title={Winning arguments: Interaction dynamics and persuasion strategies in good-faith online discussions},
  author={Tan, Chenhao and Niculae, Vlad and Danescu-Niculescu-Mizil, Cristian and Lee, Lillian},
  booktitle={Proceedings of the 25th international conference on world wide web},
  pages={613--624},
  year={2016}
}

@inproceedings{atkinson+srinivasan+tan:19,
    title = "What Gets Echoed? Understanding the {\textquotedblleft}Pointers{\textquotedblright} in Explanations of Persuasive Arguments",
    author = "Atkinson, David  and
      Srinivasan, Kumar Bhargav  and
      Tan, Chenhao",
    editor = "Inui, Kentaro  and
      Jiang, Jing  and
      Ng, Vincent  and
      Wan, Xiaojun",
    booktitle = "Proceedings of the 2019 Conference on Empirical Methods in Natural Language Processing and the 9th International Joint Conference on Natural Language Processing (EMNLP-IJCNLP)",
    month = nov,
    year = "2019",
    address = "Hong Kong, China",
    publisher = "Association for Computational Linguistics",
    url = "https://aclanthology.org/D19-1289/",
    doi = "10.18653/v1/D19-1289",
    pages = "2911--2921"
}

@article{xu2025confronting,
  title={Confronting verbalized uncertainty: Understanding how LLM’s verbalized uncertainty influences users in AI-assisted decision-making},
  author={Xu, Zhengtao and Song, Tianqi and Lee, Yi-Chieh},
  journal={International Journal of Human-Computer Studies},
  pages={103455},
  year={2025},
  publisher={Elsevier}
}

@article{li2024utilizing,
  title={Utilizing human behavior modeling to manipulate explanations in AI-assisted decision making: the good, the bad, and the scary},
  author={Li, Zhuoyan and Yin, Ming},
  journal={Advances in Neural Information Processing Systems},
  volume={37},
  pages={5025--5047},
  year={2024}
}

@article{sabour2025human,
  title={Human Decision-making is Susceptible to AI-driven Manipulation},
  author={Sabour, Sahand and Liu, June M and Liu, Siyang and Yao, Chris Z and Cui, Shiyao and Zhang, Xuanming and Zhang, Wen and Cao, Yaru and Bhat, Advait and Guan, Jian and others},
  journal={arXiv preprint arXiv:2502.07663},
  year={2025}
}

@article{zhou2024relying,
  title={Relying on the Unreliable: The Impact of Language Models' Reluctance to Express Uncertainty},
  author={Zhou, Kaitlyn and Hwang, Jena D and Ren, Xiang and Sap, Maarten},
  journal={arXiv preprint arXiv:2401.06730},
  year={2024}
}

@article{costello2024durably,
  title={Durably reducing conspiracy beliefs through dialogues with AI},
  author={Costello, Thomas H and Pennycook, Gordon and Rand, David G},
  journal={Science},
  volume={385},
  number={6714},
  pages={eadq1814},
  year={2024},
  publisher={American Association for the Advancement of Science}
}

@article{eigner2024determinants,
  title={Determinants of llm-assisted decision-making},
  author={Eigner, Eva and H{\"a}ndler, Thorsten},
  journal={arXiv preprint arXiv:2402.17385},
  year={2024}
}

@article{yan2024knownet,
  title={Knownet: Guided health information seeking from llms via knowledge graph integration},
  author={Yan, Youfu and Hou, Yu and Xiao, Yongkang and Zhang, Rui and Wang, Qianwen},
  journal={IEEE Transactions on Visualization and Computer Graphics},
  year={2024},
  publisher={IEEE}
}

@article{danry2024deceptive,
  title={Deceptive AI systems that give explanations are more convincing than honest AI systems and can amplify belief in misinformation},
  author={Danry, Valdemar and Pataranutaporn, Pat and Groh, Matthew and Epstein, Ziv and Maes, Pattie},
  journal={arXiv preprint arXiv:2408.00024},
  year={2024}
}

@article{borah2025persuasion,
  title={Persuasion at Play: Understanding Misinformation Dynamics in Demographic-Aware Human-LLM Interactions},
  author={Borah, Angana and Mihalcea, Rada and P{\'e}rez-Rosas, Ver{\'o}nica},
  journal={arXiv preprint arXiv:2503.02038},
  year={2025}
}

@article{anglin2019beliefs,
  title={Do beliefs yield to evidence? Examining belief perseverance vs. change in response to congruent empirical findings},
  author={Anglin, Stephanie M},
  journal={Journal of Experimental Social Psychology},
  volume={82},
  pages={176--199},
  year={2019},
  publisher={Elsevier}
}

@article{ji2023survey,
  title={Survey of hallucination in natural language generation},
  author={Ji, Ziwei and Lee, Nayeon and Frieske, Rita and Yu, Tiezheng and Su, Dan and Xu, Yan and Ishii, Etsuko and Bang, Ye Jin and Madotto, Andrea and Fung, Pascale},
  journal={ACM computing surveys},
  volume={55},
  number={12},
  pages={1--38},
  year={2023},
  publisher={ACM New York, NY}
}

@article{ienca2023artificial,
  title={On artificial intelligence and manipulation},
  author={Ienca, Marcello},
  journal={Topoi},
  volume={42},
  number={3},
  pages={833--842},
  year={2023},
  publisher={Springer}
}

@online{durmus2024persuasion,
author = {Esin Durmus and Liane Lovitt and Alex Tamkin and Stuart Ritchie and Jack Clark and Deep Ganguli},
title = {Measuring the Persuasiveness of Language Models},
date = {2024-04-09},
year = {2024},
url = {https://www.anthropic.com/news/measuring-model-persuasiveness},
}

@article{li2025text,
  title={From Text to Trust: Empowering AI-assisted Decision Making with Adaptive LLM-powered Analysis},
  author={Li, Zhuoyan and Zhu, Hangxiao and Lu, Zhuoran and Xiao, Ziang and Yin, Ming},
  journal={arXiv preprint arXiv:2502.11919},
  year={2025}
}

@inproceedings{kim2024m,
  title={" I'm Not Sure, But...": Examining the Impact of Large Language Models' Uncertainty Expression on User Reliance and Trust},
  author={Kim, Sunnie SY and Liao, Q Vera and Vorvoreanu, Mihaela and Ballard, Stephanie and Vaughan, Jennifer Wortman},
  booktitle={Proceedings of the 2024 ACM conference on fairness, accountability, and transparency},
  pages={822--835},
  year={2024}
}

@article{liu2024human,
  title={Human-centered NLP Fact-checking: Co-Designing with Fact-checkers using Matchmaking for AI},
  author={Liu, Houjiang and Das, Anubrata and Boltz, Alexander and Zhou, Didi and Pinaroc, Daisy and Lease, Matthew and Lee, Min Kyung},
  journal={Proceedings of the ACM on Human-Computer Interaction},
  volume={8},
  number={CSCW2},
  pages={1--44},
  year={2024},
  publisher={ACM New York, NY, USA}
}

@inproceedings{schmitt2024role,
  title={The role of explainability in collaborative human-AI disinformation detection},
  author={Schmitt, Vera and Villa-Arenas, Luis-Felipe and Feldhus, NIls and Meyer, Joachim and Spang, Robert P and M{\"o}ller, Sebastian},
  booktitle={Proceedings of the 2024 ACM conference on fairness, accountability, and transparency},
  pages={2157--2174},
  year={2024}
}

@inproceedings{danry2025deceptive,
  title={Deceptive explanations by large language models lead people to change their beliefs about misinformation more often than honest explanations},
  author={Danry, Valdemar and Pataranutaporn, Pat and Groh, Matthew and Epstein, Ziv},
  booktitle={Proceedings of the 2025 CHI Conference on Human Factors in Computing Systems},
  pages={1--31},
  year={2025}
}

@inproceedings{ding2025citations,
  title={Citations and trust in llm generated responses},
  author={Ding, Yifan and Facciani, Matthew and Joyce, Ellen and Poudel, Amrit and Bhattacharya, Sanmitra and Veeramani, Balaji and Aguinaga, Sal and Weninger, Tim},
  booktitle={Proceedings of the AAAI Conference on Artificial Intelligence},
  volume={39},
  number={22},
  pages={23787--23795},
  year={2025}
}

@article{salvi2025conversational,
  title={On the conversational persuasiveness of GPT-4},
  author={Salvi, Francesco and Horta Ribeiro, Manoel and Gallotti, Riccardo and West, Robert},
  journal={Nature Human Behaviour},
  pages={1--9},
  year={2025},
  publisher={Nature Publishing Group UK London}
}

@article{sun2025explaining,
  title={Explaining Sources of Uncertainty in Automated Fact-Checking},
  author={Sun, Jingyi and Warren, Greta and Shklovski, Irina and Augenstein, Isabelle},
  journal={arXiv preprint arXiv:2505.17855},
  year={2025}
}

@article{tao2025revisiting,
  title={Revisiting Uncertainty Estimation and Calibration of Large Language Models},
  author={Tao, Linwei and Yeh, Yi-Fan and Dong, Minjing and Huang, Tao and Torr, Philip and Xu, Chang},
  journal={arXiv preprint arXiv:2505.23854},
  year={2025}
}

@article{wang2024calibrating,
  title={Calibrating expressions of certainty},
  author={Wang, Peiqi and Lam, Barbara D and Liu, Yingcheng and Asgari-Targhi, Ameneh and Panda, Rameswar and Wells, William M and Kapur, Tina and Golland, Polina},
  journal={arXiv preprint arXiv:2410.04315},
  year={2024}
}

@String{Computing = "Computing" }

@String{Computer = "{IEEE} Computer" }

@String{Springer = "Springer-Verlag" }

\end{document}